\documentclass[a4paper,fleqn,usenatbib]{mnras}

\pdfoutput=1

\usepackage[T1]{fontenc}
\usepackage{ae,aecompl}

\usepackage{graphicx}	
\usepackage{amsmath}	
\usepackage{amssymb}	
\usepackage{ulem}       
\usepackage{cancel}     

\title[TDE Accretion Disks]{GRRMHD Simulations of Tidal Disruption Event Accretion Disks around Supermassive Black Holes:\\ Jet Formation, Spectra, and Detectability}

\author[Brandon Curd \& Ramesh Narayan]{
Brandon Curd,$^{1}$\thanks{E-mail: brandon.curd@cfa.harvard.edu}
Ramesh Narayan,$^{1}$
\\
$^{1}$ Harvard-Smithsonian Center for Astrophysics, 60 Garden Street, Cambridge, MA 02138, USA
}

\date{Accepted XXX. Received YYY; in original form ZZZ}

\pubyear{2018}

\begin{document}
\label{firstpage}
\pagerange{\pageref{firstpage}--\pageref{lastpage}}
\maketitle

\begin{abstract}
We report results from general relativistic radiation MHD (GRRMHD) simulations of a super-Eddington black hole (BH) accretion disk formed as a result of a tidal disruption event (TDE). We consider the fiducial case of a solar mass star on a mildly penetrating orbit disrupted by a supermassive BH of mass $10^6 \, M_\odot$, and consider the epoch of peak fall back rate. We post-process the simulation data to compute viewing angle dependent spectra. We perform a parameter study of the dynamics of the accretion disk as a function of BH spin and magnetic flux, and compute model spectra as a function of the viewing angle of the observer. We also consider detection limits based on the model spectra. We find that an accretion disk with a relatively weak magnetic field around the BH (so-called SANE regime of accretion) does not launch a relativistic jet, whether or not the BH is rotating. Such models reasonably reproduce several observational properties of non-jetted TDEs. The same is also true for a non-rotating BH with a strong magnetic field (MAD regime). One of our simulations has a rapidly rotating BH (spin parameter 0.9) as well as a MAD accretion disk. This model launches a powerful relativistic jet, which is powered by the BH spin energy. It reproduces the high energy emission and jet structure of the jetted TDE \textit{Swift} J1644+57 surprisingly well. Jetted TDEs  may thus correspond to the subset of TDE systems that have both a rapidly spinning BH and MAD accretion.
\end{abstract}




\section{Introduction}
When a star wanders too close to the black hole (BH) at the center of its galaxy, the tidal gravitational forces acting on the star overcome its self-gravity and ultimately disrupt the star \citep{Hills1975,Rees1988,Phinney1989,Evans1989}. The disruption leads to streams of bound and unbound material, with roughly half of the disrupted mass returning to form an accretion disk. The rate at which the bound material returns, or the ``fall back rate'', declines with time and scales roughly as $\dot{M}_{\rm{fb}}\propto t^{-5/3}$. The energy released from fall back and accretion is predicted to produce a transient that peaks in the UV and soft X-rays \citep{Cannizzo1990}. The emission is expected to decline in a similar fashion to the fall back rate.

Several decades after the initial theoretical bedrock was laid in the above-cited papers, several TDEs were detected by the soft X-ray telescope \textit{ROSAT} \citep{Komossa2015}. These TDEs peaked in the soft X-ray with $L_X \lesssim 10^{44}$ ${\rm erg\,s^{-1}}$ within $\sim$months, and followed a roughly $t^{-5/3}$ decay as predicted, fading over $\sim$years. The spectra appeared very soft at peak and hardened on the timescale of a few years.

More recently, UV and optical wide-field surveys revealed TDEs flaring in the UV/optical, with several of these events also showing an X-ray flare \citep{Gezari2006,Gezari2008,Gezari2009,Komossa2008,vanVelzen2011,Holoien2016a}. Where both UV/optical and X-ray data are available, it appears that the two components are associated with different regions, with the UV/optical emission having a characteristic temperature $T\sim 10^4$ K and the X-ray component $T\sim 10^5$. It is  not clear if the spectrum is simply the sum of two thermal components with these temperatures, as there are no observations of the FUV component. 

TDEs are capable of launching powerful relativistic jets, as the \textit{Swift} J1644+57 (J1644 hereafter) event showed \citep{Bloom2011,Burrows2011,Zauderer2011}. This transient, which is the prototypical example of what is referred to as a ``jetted TDE,'' was initially detected as a gamma-ray burst (GRB) in a quiescent host galaxy; however, the light curve was rapidly variable and remained visible for much longer than GRBs do. The peak X-ray luminosity of $L_X \sim 10^{48}$ ${\rm erg\,s^{-1}}$ was much higher than in TDEs previously detected. Follow-up radio observations revealed radio emission from a relativistic jet shocking with the surrounding medium. The emission was likely the result of a jet with $\Gamma \gtrsim 10$, viewed near the jet axis \citep{Metzger2012}. Two other jetted TDE candidates, \textit{Swift} J2058+0516 (J2058 hereafter) and \textit{Swift} J1112-8238 (J1112 hereafter), have since been detected \citep{Cenko2012,Brown2015}.

Recent theoretical works have investigated the hydrodynamics \citep{Ramirez-Ruiz2009,Guillochon2013,Shiokawa2015,Bonnerot2016a,Hayasaki2016}, emission properties \citep{Strubbe2009,Lodato2011,Guillochon2014}, impact of BH spin \citep{Kesden2012a,Kesden2012b,Stone2012,Guillochon2015}, and jet properties \citep{Giannios2011,Piran2015,Lu2017}, of TDEs. A fundamental question regarding the nature of TDEs is the mass accretion rate onto the BH. The possibility of super-Eddington rates was pointed out early on \citep{Rees1988}, but this requires that the circularization and viscous dissipation timescales are small relative to the fall back time of the most bound material. \citet{Mockler2018} find that the light curves of many observed TDEs are consistent with a short viscous dissipation timescale; however, numerical simulations suggest circularization times that are several times longer than the fall back time \citep{Bonnerot2016a,Hayasaki2016}. 

Super-Eddington, rotating black holes produce highly relativistic jets \citep{Sadowski2015b, Narayan2017, Riordan2017}, fueled at least in part by the extraction of spin energy from the black hole via the Blandford-Znajek process \citep{Blandford1977}. A magnetically arrested accretion disk (MAD, see \citealt{Narayan2003}, also \citealt{BisnovatyiKogan1974, Igumenshchev2003, Tchekhovskoy2014}) provides an attractive explanation for jetted TDEs since magnetohydrodynamical (MHD) simulations have shown they can launch powerful jets, e.g. \citet{Tchekhovskoy2011}. Whether the magnetic field needed for the Blandford-Znajek mechanism to operate is from the disrupted star itself or was already present in the form of a fossil disk is still an open question. \citet{Kelley2014} demonstrated that accumulating a magnetic flux sufficient to launch a jet is possible if the BH had a fossil accretion disk threaded with a magnetic field. They found that the tidal stream from the disrupted star effectively drags the remnant magnetic field in. Studies of the magnetic field evolution during a TDE suggest that field amplification during disruption may also be sufficient to provide the magnetic flux \citep{Guillochon2017,Bonnerot2017}. 

Observations indicate that there is a zoo of TDE properties despite the somewhat simple physics. Some TDEs are seen only in the UV/optical while others have been detected in both soft X-ray and UV/optical bands. The rarity of jetted TDEs suggests that there is a specific region of parameter space where a jet will form. It is reasonable to suppose that this range of behavior is related to the viewing angle as well as the properties of the BH and star.

\citet{Dai2018} examined the viewing angle dependence. They presented the first 3D general relativistic radiation magnetohydrodynamics (GRRMHD hereafter) simulation of a TDE accretion disk near the peak fall back accretion rate. They investigated the case of a modestly rotating BH of mass $5\times10^6\,M_\odot$ with spin parameter, $a_* \equiv a/M = 0.8$, and mean accretion rate $\langle \dot{M} \rangle \sim 15\,\dot{M}_{\rm{Edd}}$. Their model led to a mildly relativistic, wide angle, ultrafast outflow with a Lorentz factor $\Gamma \sim 1.4$. The viewing angle-dependent spectra computed from their model suggests that the optical to X-ray flux ratio increases as the observer moves towards the disk plane. While their model may only apply to ``non-jetted TDEs,'' i.e., TDEs that do not launch an ultra relativistic jet (though they may have a mildly relativistic jet), it does offer a qualitative picture of how the viewing angle of the observer can affect the observed properties of a TDE.

In the present work, we use GRRMHD simulations in 2D and 3D to study the post-fallback accretion disk for the fiducial case of a $10^6 \, M_\odot$ supermassive black hole (SMBH) that disrupts a solar mass star on a slightly penetrating orbit. We run four models to investigate the properties of super-Eddington accreting TDEs as well as the conditions for the launching of an ultra relativistic jet. We consider two values of the BH spin: $a_*=0,\, 0.9$. For each spin, we initialize the magnetic field so that the BH either does or does not build up a high magnetic flux. These four models correspond to a TDE accretion disk around a low (high) spin BH which has (has not) become a magnetically arrested disk (MAD). We post-process the simulation output using a fully general relativistic radiative transfer code. We study the spectra from each model as a function of viewing angle and investigate how BH spin, magnetic field strength, viewing angle and interstellar extinction change the properties of the observed TDE emission. The angular momentum of the initial disk is aligned with the BH spin in the models we consider in this work. If they were misaligned, Lense-Thirring torques would cause the disk to precess \citep{Fragile2007}. In addition, jet precession would occur until a MAD state was achieved, after which wobbling during jet alignment with the BH spin might cause intense flaring \citep{Tchekhovskoy2014}. A larger amount of mass may also be ejected in this scenario as the precessing jet sweeps up material in its path.

In Section 2, we review the physics involved in a TDE and discuss our choice of parameters for initial conditions. In Section 3, we describe our numerical methods. In Section 4, we describe the results of our simulations in terms of the dynamics and energetics. In Section 5, we discuss the viewing angle dependent spectra computed from the four models and describe the emission in detail. We then compute relevant properties and compare our models to observations of jetted and non-jetted TDEs. We conclude in Section 6.

\section{Tidal Disruption Event Physics}

A star will be disrupted by a supermassive BH (SMBH) if it comes closer than the tidal radius \citep{Hills1975}:
\begin{equation} \label{eq:Rt}
  R_t \sim 7\times 10^{12}\, m_6^{1/3}m_*^{-1/3}r_* \,\rm{cm},
\end{equation}
where $m_6=M_{\rm{BH}}/10^6\,M_\odot$ is the dimensionless mass of the SMBH, $m_*=M_{\rm{*}}/M_\odot$ is the dimensionless mass of the disrupted star, and $r_*=R_{\rm{*}}/R_\odot$ is its dimensionless radius. It is useful to describe the disruption in terms of the ratio between the tidal radius and pericenter separation of the star, $\beta = R_t/R_p$. Disruption occurs for $1 \lesssim \beta \lesssim R_t/R_S$, where $R_S = 2GM_{\rm{BH}}/c^2$ is the Schwarzschild radius of the BH.

Following pericentric passage, nearly half of the material in the disrupted star remains bound and will return to form a disk, while the other half is unbound and escapes \citep{Rees1988}. The spread in specific orbital energy of the streams is effectively `frozen in' at the tidal radius and has a spread given by \citep{Stone2013}
\begin{equation} \label{eq:depsilon}
  \Delta\epsilon = \dfrac{G M_{\rm{BH}}R_*}{R_t^2}.
\end{equation} 
The bound material returns at the fall back rate, which initially peaks at a super-Eddington rate for $M_{\rm{BH}} \lesssim 3\times 10^7 \,M_\odot$, and then falls off as $\dot{M}_{\rm{fb}} \propto t^{-5/3}$. More recent works show that, depending on the stellar properties, the fall back rate can deviate at early times from the $t^{-5/3}$ decline \citep{Lodato2009}. For $m_6\sim1$, the peak fall back accretion rate is expected to be roughly $\sim 100 \dot{M}_{\rm{Edd}}$ \citep{Stone2013}.

\citet{Rees1988} suggested that internal dissipation would lead the bound material to form a radiation-supported torus with a density maximum at $\sim\!\!\! R_t$. This disk is fed by the accretion of the bound material in the tidal stream. Hydrodynamical simulations \citep{Shiokawa2015,Bonnerot2016a,Hayasaki2016,Sadowski2016a} and semi-analytic studies \citep{Dai2013,Dai2015,Guillochon2015} have demonstrated that the formation of a thick torus is possible within a certain parameter space. For highly penetrating orbits, stream-stream collisions can drive disk formation on a relatively short timescale. The rate at which material actually accretes onto the SMBH from the torus is then mediated by the viscous dissipation timescale. If the viscous dissipation timescale is short, the mass accretion rate is expected to closely match the fall back accretion rate.

In this work, we simulate the accretion flow of an efficiently circularizing TDE disk for the case of a mildly penetrating disruption $(\beta \sim 2.5)$ of a Sun-like star of mass $M_* = M_\odot$ and solar metallicity. The disrupting hole is a $10^6 \, M_\odot$ SMBH. For such an event, the star is initially on a parabolic trajectory and the circularization radius of the disk can be approximated as
\begin{equation} \label{eq:Rc}
  R_c = 2R_p = 2R_t/\beta. 
\end{equation} 

In initializing the gas in the simulations presented in this work, we set the Bernoulli number to the binding energy as given in equation (\ref{eq:depsilon}). Following \citet{Rees1988}, we assume that the debris that forms the disk all comes in with roughly the same specific angular momentum and that the density maximum of the resulting disk occurs at $R_c$. We also assume that the viscous dissipation timescale is short enough that the mass accretion rate onto the black hole is similar to the fall back rate.

\section{Numerical Methods}

\subsection{\textsc{KORAL}}
The GRRMHD simulations presented in this work were performed using the code \textsc{KORAL} \citep{Sadowski2013a,Sadowski2014,Sadowski2015,Sadowski2017}, which solves the following conservation equations of MHD in a fixed, arbitrary spacetime using finite-difference methods:
\begin{align}
  (\rho u^\mu)_{;\mu} &= 0, \label{eq:drhou} \\
  (T^\mu_{\ \, \nu})_{;\mu} &= G_\nu, \label{eq:dTmunu} \\
  (R^\mu_{\ \, \nu})_{;\mu} &= -G_\nu, \label{eq:dRmunu} \\
  (n u_R^\mu)_{;\mu} &= \dot{n}, \label{eq:dn} 
\end{align}
where we use standard relativistic notation, with a semicolon denoting a covariant derivative ($\nabla_\mu V^\nu \equiv V^\nu_{\ \, ;\mu}$) and a dot denoting a time derivative in  the comoving frame ($\dot{n} \equiv dn/d\tau$, with $\tau$ being the proper time of the relevant fluid). Here, $\rho$ is the gas density in the comoving fluid frame; $u^\mu$ are the components of the gas four-velocity as measured in the ``lab frame''; $u_R^\mu$ are the components of the radiation fluid four-velocity (see \citealt{Sadowski2015b}); $T^\mu_{\ \, \nu}$ is the MHD stress-energy tensor in the lab frame,
\begin{equation} \label{eq:Tmunu}
  T^\mu_{\ \, \nu} = (\rho + u_g+ p_g + b^2)u^\mu u_\nu + \left(p_g + \dfrac{1}{2}b^2 \right)\delta^\mu_{\ \, \nu} - b^\mu b_\nu;
\end{equation}
$R^\mu_{\ \, \nu}$ is the stress-energy tensor of radiation \citep{Sadowski2015b}; $G_\nu$ is the radiative four-force which describes the interaction between gas and radiation \citep{Sadowski2014}; and $n$ is the photon number density. The internal energy $u_g$ and gas pressure $p_g$ are related by $p_g=(\gamma - 1)u_g$, where $\gamma$ is the adiabatic index of the gas, and the magnetic field four-vector $b^\mu$ is evolved following the ideal MHD induction equation \citep{Gammie2003}. 

The radiative stress-energy tensor is obtained from the evolved radiative primitives, namely, the four velocity of the radiative rest frame and the radiative energy density $\hat{E}$ in this frame \citep{Sadowski2013a,Sadowski2014}. The M1 closure scheme \citep{Levermore1984} is used, modified by the addition of radiative viscosity \citep{Sadowski2015}.

The interaction between gas and radiation (absorption, emission, and scattering) is described by the radiation four-force $G_\nu$. The opposite signs of this quantity in the conservation equations for gas and radiation stress-energy reflect the fact that the gas-radiation interaction is conservative, i.e. energy and momentum lost by one is gained by the other. For a detailed description of the four-force see \citet{Sadowski2017}.

Two of the simulations described in this work were performed in 2D, assuming axisymmetry and using the mean-field dynamo model described in \citet{Sadowski2015}. These 2D runs correspond to the SANE (``Standard and Normal Evolution'', see \citealt{Narayan2012}) regime of accretion, where the magnetic field that accumulates around the BH is relatively weak. Previous work has shown that, for SANE simulations, 2D and 3D runs give similar results \citep{Sadowski2016b, Narayan2017}. The remaining
two simulations correspond to the MAD regime, and these require 3D. For these, we use a similar grid to the 2D simulations, but with lower resolution. We initially ran the simulations in 2D with the mean-field dynamo  until the accretion flow became MAD. We then switched off the mean-field dynamo, re-gridded the data on a 3D grid by copying the primitives in azimuth and perturbing the azimuthal fluid velocity, and then continued evolving the simulation in 3D. We find that the effects of the initial 2D run and the regridding dissipate completely within $5000\,t_g$ after switching to 3D, where $t_g = GM_{\rm BH}/c^3$ is the gravitational time.

For both 2D and 3D simulations we use modified Kerr-Schild coordinates with the inner edge of the domain inside the BH horizon. The radial grid cells are spaced logarithmically in radius and the cells in polar angle $\theta$ are smaller towards the equatorial plane. The cells are equally spaced in azimuth, covering a range of $\pi$. At the inner and outer radial boundaries, we use outflow boundary conditions, which prevent the inflow of gas and radiation. At the polar boundaries, we use a reflective boundary condition, while in the azimuthal direction we employ a periodic boundary condition. Table \ref{tab:tab1} lists the resolution, grid parameters and other relevant information of the four simulations. The models are given descriptive names -- \texttt{s00}, \texttt{m00}, \texttt{s09}, \texttt{m09} -- which distinguish whether they are SANE or MAD, and identify the BH spin.

We have verified that the fastest growing mode of the magnetorotational instability (MRI, \citealt{Balbus1991}) is adequately resolved within each simulation. For this we compute the quantities \citep{Hawley2011},
\begin{align}
  Q_\theta = \dfrac{2\pi}{\Omega\, dx^\theta}\dfrac{|b^\theta|}{\sqrt{4\pi\rho}}, \label{eq:Qtheta} \\
  Q_\phi = \dfrac{2\pi}{\Omega\, dx^\phi}\dfrac{|b^\phi|}{\sqrt{4\pi\rho}}, \label{eq:Qphi}
\end{align}
where $dx^i$ (the grid cell size) and $b^i$ (the magnetic field strength) are both evaluated in the orthonormal frame, $\Omega$ is the angular velocity, and $\rho$ is the gas density. For the 2D SANE models, we find $Q_\theta \sim 50$ in the disk region within $r=100\, r_g$, where $r_g = GM_{\rm BH}/c^2$ is the gravitational radius. For the 3D MAD models, we find $Q_\theta \sim 50$ and $Q_\phi \sim 20$ within the same region. In both cases, this is sufficient to resolve the MRI \citep{Hawley2011}.

\subsection{\textsc{HEROIC}} \label{sec:HEROIC}
The output from each GRRMHD simulation was post-processed using the general relativistic multi-dimensional radiative transfer code \textsc{HEROIC} \citep{Zhu2015, Narayan2016}, using the procedures described in \citet{Narayan2017}. In brief, time- and azimuth-averaged output from the simulation is transferred to an axisymmetric grid with uniform spacing in $\log r$ and $\theta$. The radiation field in each cell within this grid is described by rays oriented along 162 angular directions distributed uniformly on the sphere. The intensity distribution of each ray is described by a spectrum with 160 frequencies spaced uniformly in $\log \nu$ from $\nu=10^8$\,Hz to $10^{24}$\,Hz.

The global radiation field over the entire grid is obtained by solving the radiative transfer equation iteratively. \textsc{HEROIC} uses the same continuum opacities as those in \textsc{KORAL}, viz., free-free, thermal synchrotron, atomic processes (via the model of \citealt{Sutherland_1993}), and Comptonization of all of these. However, \textsc{HEROIC} includes the frequency dependence of the various opacities (only approximately in the case of atomic edges), whereas \textsc{KORAL} works with gray opacities. In the process of solving for the radiation field, \textsc{HEROIC} also solves for the gas temperature in selected regions of the grid, using the viscous heating rate estimated in \textsc{KORAL} and applying the condition of energy balance. The temperature solution is restricted to those regions of the flow that either have reached steady state in the GRRMHD simulation or from which radiation can diffuse out in less than the duration of the simulation ($20000t_g$ or $25000t_g$, see Table \ref{tab:tab1}). For the remaining cells, we keep the temperature fixed at the value obtained from the GRRMHD simulation. In practice, the only cells with fixed temperature are those deep inside the torus in regions of large optical depth. All regions from which any significant radiation reaches an external observer have their temperature solved for self-consistently.

After we obtain a converged solution for the radiation field from \textsc{HEROIC}, we use general relativistic ray-tracing to calculate the radiation spectrum seen by observers located at different viewing angles. We also compute synthetic  images of the accretion flow to identify which regions of the flow contribute to which parts of the spectrum.

\section{Simulations}

\subsection{Units}

We define the gravitational radius $r_g$ and the gravitational time $t_g$ by
\begin{equation} \label{eq:rgtg}
r_g = \frac{GM_{\rm BH}}{c^2}, \qquad t_g = \frac{GM_{\rm BH}}{c^3}.
\end{equation}
We use these as our units of length and time. Often, we set $G=c=1$, so the above relations would be equivalent to $r_g=t_g=M_{\rm BH}$. However, we occasionally restore $G$ and $c$ when we feel it helps to keep track of units.

We adopt the following definition for the Eddington mass accretion rate:
\begin{equation} \label{eq:mdotEdd}
  \dot{M}_{\rm{Edd}} = \dfrac{L_{\rm{Edd}}}{\eta c^2},
\end{equation}
where $L_{\rm{Edd}} = 1.25\times 10^{38}\, (M/M_\odot)\, {\rm erg\,s^{-1}}$ is the Eddington luminosity, $\eta$ is the radiative efficiency of a thin disk around a BH with spin parameter $a_*$,
\begin{equation} \label{eq:etaNT}
  \eta = 1 - \sqrt{1 - \dfrac{2}{3 r_{\rm{ISCO}}}},
\end{equation}
and $r_{\rm{ISCO}}=3+Z_2 - \sqrt{(3-Z_1)(3+Z_1+2Z_2)}$ is the radius of the Innermost Stable Circular Orbit (ISCO, \citealt{Novikov1973}) in the Kerr metric, where $Z_1 = 1 + (1-a_*^2)^{1/3}\left((1+a_*)^{1/3}+(1-a_*)^{1/3}\right)$ and $Z_2 = \sqrt{3a_*^2 + Z_1^2}$.

\subsection{Definitions}

\subsubsection{Useful Quantities}
In a quasi-steady state, the net accretion rate is constant in time and independent of radius. The accretion rate can be estimated from the simulation data by computing the following integral at any radius within the region of inflow equilibrium:
\begin{equation} \label{eq:mdot}
  \dot{M} = -\int_0^\pi \int_0^{2\pi} \sqrt{-g}\rho u^r d\phi d\theta.
\end{equation}
Here $u^r$ is the radial component of the four-velocity and $g$ is the determinant of the metric (which is proportional to $r^4$). Positive values of $\rho u^r$ correspond to an outward mass flux, hence the minus sign in front of the integral. In this paper, we compute $\dot{M}$ at the event horizon, which is located at $r_H = 1+\sqrt{1-a_*^2}$.

The flow of energy in different forms is fundamental to understanding the properties of GRRMHD simulations. The most fundamental quantity is the total luminosity in all forms of energy (radiative, electromagnetic, thermal, gravitational, and binding energy) minus the rest mass energy of the accreted gas. We call this the \textit{total luminosity},
\begin{equation} \label{eq:Ltot}
  L_{\rm{tot}} = -\int_{\theta_{\rm{min}}}^{\theta_{\rm{max}}}\int_0^\pi\sqrt{-g} (T^r_{\ \, t} + R^r_{\ \, t} + \rho u^r) d\phi d\theta,
\end{equation}
where we integrate the radial flux of energy carried by gas plus magnetic field ($T^r_{\ \, t}$) and radiation ($R^r_{\ \, t}$), and subtract out the rest-mass energy ($\rho u^r$, since it does not lead to observational consequences for an observer at infinity). Note that positive values of $T^r_{\ \, t}$ and $R^r_{\ \, t}$ correspond to an inward flux of energy, hence the minus sign in front of the integral. In a stationary state, the total luminosity integrated over all angles is independent of radius. It gives the total luminosity of the whole system, and represents for instance the luminosity, both radiative and kinetic, seen at infinity. 

It is also useful to define individual components of the energy flow (e.g., \citealt{Sadowski2016c}). The corresponding luminosities will not
be independent of radius since energy could be transferred from one form to another as gas flows inward or outward. Nevertheless, these luminosities often provide useful insights. The \textit{radiative luminosity} is given by:
\begin{equation} \label{eq:Lrad}
  L_{\rm{rad}} = -\int_{\theta_{\rm{min}}}^{\theta_{\rm{max}}}\int_0^{2\pi}\sqrt{-g} R^r_{\ \, t} d\phi d\theta,
\end{equation}
which gives the flux of radiation energy through a surface at a given radius. 
A related luminosity we consider is the radiative isotropic equivalent luminosity, which is the radiative luminosity of the source if we take the radiative flux corresponding to a given direction $\theta$ (in the case of 3D models, we average over $\phi$) and assume that the source emits the same flux isotropically over all angles:
\begin{equation} \label{eq:Liso}
  L_{\rm{iso}}(r,\theta) = 4\pi r^2 F_{\rm{rad}}(r,\theta).
\end{equation}
In addition to these radiative luminosities, we similarly define the kinetic luminosity, magnetic (i.e., Poynting) luminosity, etc.

The relativistic kinetic energy density is given by $\epsilon_k = \rho (u^t - 1)$. The total kinetic energy within any given region of the grid, e.g., between $r_{\rm min}$ and $r_{\rm max}$ and between $\theta_{\rm min}$ and $\theta_{\rm max}$, is found by integrating over the corresponding volume. This quantity is given by:
\begin{equation} \label{eq:Ek}
  E_k = \int_{r_{\rm{min}}}^{r_{\rm{max}}}\int_{\theta_{\rm{min}}}^{\theta_{\rm{max}}}\int_0^{2\pi} \sqrt{-g} \rho \left(u^t - 1\right) d\phi d\theta dr,
\end{equation}
and is sometimes useful when analyzing physics in the jet.

Another important quantity is the efficiency, which is the fraction of the accreted rest-mass energy that is converted into any particular form of energy. For any given energy form, we define the efficiency via the corresponding luminosity:
\begin{equation} \label{eq:etai}
  \eta_i = \dfrac{L_i}{\dot{M}c^2},
\end{equation}
where $L_i$ is any of the luminosities above. When studying efficiencies, a useful benchmark is the radiative efficiency of a general relativistic thin disk \citep{Novikov1973}, which is $\eta_{\rm{NT}}=0.0572$ for $a_*=0$ and $\eta_{\rm{NT}}=0.1558$ for $a_*=0.9$ (the two spins we consider in this paper).

We quantify the magnetic field strength at the BH horizon through the dimensionless magnetic flux parameter \citep{Tchekhovskoy2011}:
\begin{equation} \label{eq:phiBH}
  \Phi_{\rm{BH}}(t) = \dfrac{1}{2\sqrt{\dot{M}r_g^2 c}} \int_{\theta_{\rm{min}}}^{\theta_{\rm{max}}}\int_0^{2\pi} \sqrt{-g} \, |B^r(r_H,t)| \, d\phi d\theta,
\end{equation}
where $B^r$ is the radial component of the magnetic field and the integral is computed at the radius $r_H$ of the horizon. For geometrically thick disks such as those considered in this work, the MAD state is achieved once $\Phi_{\rm{BH}}\sim 40-50$ \citep{Tchekhovskoy2011,Tchekhovskoy2012}.

We have written the upper and lower bounds of $\theta$ in many of the integrals above as $\theta_{\rm{min}}$, $\theta_{\rm{max}}$, to signify that we do not always integrate over the entire sphere when considering these quantities. For instance, we sometimes perform the integrals over the disk, wind, and jet regions separately in order to determine where most of the energy released by the system is deposited.

We estimate the electron scattering photosphere location for an observer at infinity along the direction $(\theta,\phi)$ by integrating the optical depth radially inward from the outer boundary of the grid. Far from the BH, the curvature of spacetime is negligible, so we simply integrate at constant $(\theta,\phi)$ in the ``lab frame'' (i.e., we ignore frame-dragging in this computation):
\begin{equation} \label{eq:taues}
  \tau_{\rm{es}}(r) = \int_r^{r_{\rm{max}}} \dfrac{\rho \kappa_{\rm{es}}}{c} \left(u^t - u^r\right)\sqrt{g_{rr}}dr',
\end{equation}
where $\kappa_{\rm{es}} = 0.2(1+X)\kappa_{\rm{KN}}\,{\rm cm^2}$ is the electron scattering opacity, $\kappa_{\rm{KN}}$  is the Klein-Nishina correction factor for thermal electrons \citep{Sadowski2017}, and $r_{\rm{max}}$ is the radius corresponding to the outer boundary of the grid. For the gas and radiation temperatures in the simulations presented here, the Klein-Nishina correction is negligible and the electron scattering opacity is essentially the Thomson opacity. In this work, we choose the location of the photosphere as the $\tau_{\rm{es}}=1$ surface.

As our simulations couple gas and radiation, it is useful to consider the temperatures of the two separately. The gas temperature $T_g$ quantifies the thermal energy of the gas while the radiation temperature $T_r$ is related to the mean photon energy in a cell. As an example, if the emission of a cell were strictly black-body emission in LTE, then one would expect $T_g = T_r$. However, in regions where synchrotron emission or Compton/inverse Compton scattering are significant or the optical depth is not large, the radiation tends to have $T_r \neq T_g$. 

\subsubsection{`Jet', `Wind', and `Disk'}
For optically thick accretion flows in TDEs, one can generally identify three distinct regions. Near the BH at equatorial angles, there is an accretion disk which consists of high density gas flowing in. This transitions on the outside to a bound torus with low binding energy that serves as a gas reservoir for mass accretion. In our simulations, the torus also provides the magnetic field to trigger the MRI and feed gas onto the BH. All this constitutes the `disk'.

At more polar angles, gas and radiation flow out of the system. The outflow may be divided into (i) a `jet' with a large energy flux and high velocity, and (ii) a `wind' with a large mass flux and slower velocity. The exact separation between these two regions is somewhat ambiguous. We use a similar method to that employed in \citet{Sadowski2013b}, viz., in terms of the Bernoulli number, which is defined as follows:
\begin{equation} \label{eq:Be}
  {\rm{Be}} = -\dfrac{T^t_{\ \, t} + R^t_{\ \, t} + \rho u^t}{\rho u^t}.
\end{equation}

We use the value of Be to divide the various regions in the simulation. For the disk/torus, we simply use the condition ${\rm Be}<0$, i.e., the gas is bound to the BH. The jet and wind are both unbound with ${\rm Be} \geq 0$, and we select a critical value of ${\rm Be}$ to define the boundary between the fast jet and the slower wind. Here we adopt ${\rm{Be}}_{\rm{crit}}=0.05$, which corresponds to a particle velocity of $v/c \sim 0.3$ at infinity. We define the `jet' as those regions with ${\rm Be} \ge {\rm{Be}}_{\rm{crit}}$ and the `wind' as the regions with $0 < {\rm Be} < {\rm{Be}}_{\rm{crit}}$.

\subsection{Initial Setup}

\begin{table*}
  \centering
  \begin{tabular}{l c c c c}
    \hline
    \hline
     & \texttt{s00} (SANE, 2D) & \texttt{m00} (MAD, 3D) & \texttt{s09} (SANE, 2D) & \texttt{m09} (MAD, 3D) \\
    \hline
    $M_{\rm{BH}}/M_\odot$ & $10^6$ & $10^6$ & $10^6$ & $10^6$ \\
    $\dot{M}/\dot{M}_{\rm{Edd}}$ & $32$ & $95$ & $50$ & $150$ \\
    $L_{\rm{rad}} / L_{\rm{Edd}}$ & 3.4 & 2.2 & 4.0 & 615 \\
    $a_*$ & $0$ & $0$ & $0.9$ & $0.9$ \\
    $\Phi_{\rm{BH}}$ & $6$ & $50$ & $9$ & $51$ \\
    $\eta_r$ & 0.6\% & 0.132\% & 1.2\% & 64.0\% \\
    $\eta_t$ & 3.3\% & 1.8\% & 7.2\%  & 105.2\% \\
    $\eta_{t,\rm{jet}}/\eta_{t,\rm{wind}}$ & 1.26 & 0.17 & 2.29  & 135.59 \\
    $N_r \times N_\theta \times N_\phi$ & $512\times 320 \times 1$ & $320 \times 192 \times 32$ & $512 \times 320 \times 1$ & $320 \times 192 \times 32$ \\
    $r_{\rm{min}}/r_{\rm{max}}$ & $1.8/10^5$ & $1.8/10^5$ & $1.3/10^5$ & $1.3/10^5$ \\
    $t_{\rm{max}}$ & 20,000 & 25,000 & 20,000 & 25,000 \\
    \hline
  \end{tabular}
  \caption{Simulation parameters and properties of the four \textsc{KORAL} simulations.}
  \label{tab:tab1}
\end{table*}

We initialize the simulations with the hydrostatic equilibrium torus model of \citet{Kato2004}. Since the tidal fall back stream comes in with nearly constant specific angular momentum \citep{Rees1988}, we initialize the torus such that the fluid has the same  specific angular momentum throughout. 

The Bernoulli number of this model, assuming a constant angular momentum torus in hydrostatic equilibrium, is also constant throughout. As such, we find it convenient to set the Bernoulli number of the torus equal to the binding energy of the most bound material, which is given by equation (\ref{eq:depsilon}). This leads to a tenuously bound torus which is nearly spherical in structure with an evacuated polar region. The pressure of the gas (as determined from the model) is then redistributed between thermal gas and blackbody radiation such that local thermal equilibrium is maintained ($T_g=T_r$). The resulting torus is close to equilibrium. Appendix \ref{sec:appA} gives a more detailed description of the torus model and how it is initialized in \textsc{KORAL}.

The initial torus in the simulations described in this paper (the same torus was used for all four models) has a total mass of $0.17\,M_\odot$, which is slightly lower than the $\sim 0.5 M_\odot$ expected for the disruption of a solar mass star. The majority of TDEs are expected to be of low mass stars \citep{Kroupa1993}, so the initial disk mass is probably quite reasonable. The fall back accretion rate would still be super Eddington, as in the simulations presented herein, since the peak fall back accretion rate scales (approximately) only as the square root of the mass of the disrupted star (see \citealt{Stone2013}).

We thread the torus with a weak magnetic field whose strength is scaled such that the minimum magnetic pressure ratio:
\begin{equation} \label{eq:betam}
  \beta_m \equiv (p_{\rm{gas}} + p_r) / p_m,
\end{equation}
is $\sim 30$, where $p_{\rm{gas}}$ is the gas pressure, $p_r$ is the radiation pressure, and $p_m$ is the magnetic pressure. 
The initial radiation pressure ratio (defined in an analogous manner to $\beta_m$):
\begin{equation} \label{eq:betar}
  \beta_r \equiv (p_{\rm{gas}} + p_m) / p_r,
\end{equation}
is quite small, meaning that the torus is highly radiation dominated.

To study the conditions under which a TDE will or will not result in a jet, we perform four accretion disk simulations: two BH spins, two magnetic field geometries. The properties of the simulations are summarized in Table \ref{tab:tab1}. The two SANE models are initialized with multiple poloidal loops of magnetic field of changing polarity, while the two MAD models are initialized with a single large scale dipolar field loop. Appendix A gives details. The initial field configuration in the SANE models is designed such that relatively little magnetic flux accumulates around the BH. On the other hand, the field configuration in the MAD models ensures that the magnetic flux very rapidly builds up around the BH; the back reaction of this field causes the accretion flow to be magnetically arrested and to settle down to the MAD state \citep{Narayan2003}.

As we show below, three of the simulations, viz., \texttt{s00, s09, m00}, resemble typical non-jetted TDEs. We consider all three to be viable models of non-jetted TDEs. In the main paper, we present detailed results for one of the models, \texttt{s00}; results for the other two are summarized in Appendix C. The fourth model, \texttt{m09}, strongly resembles jetted TDEs, especially J1644. We consider this to be an excellent prototype for a jetted TDE.

\subsection{Properties of Models}

\begin{figure*}
	\includegraphics[width=\textwidth]{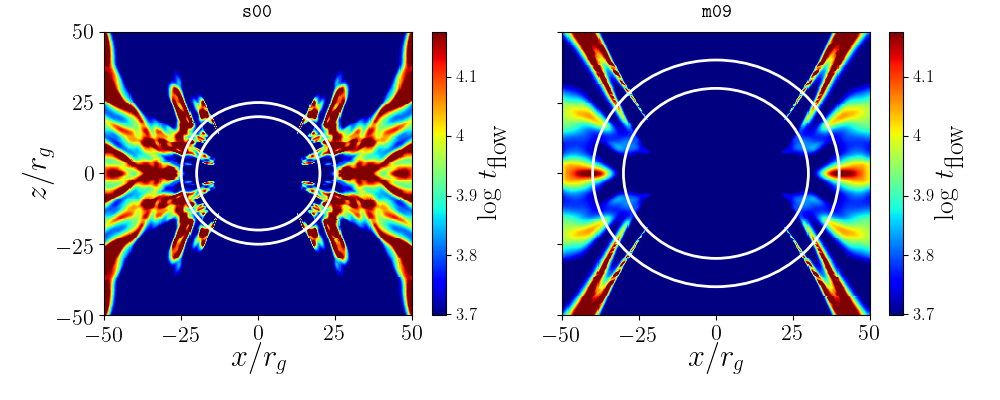}
   	\includegraphics[width=\textwidth]{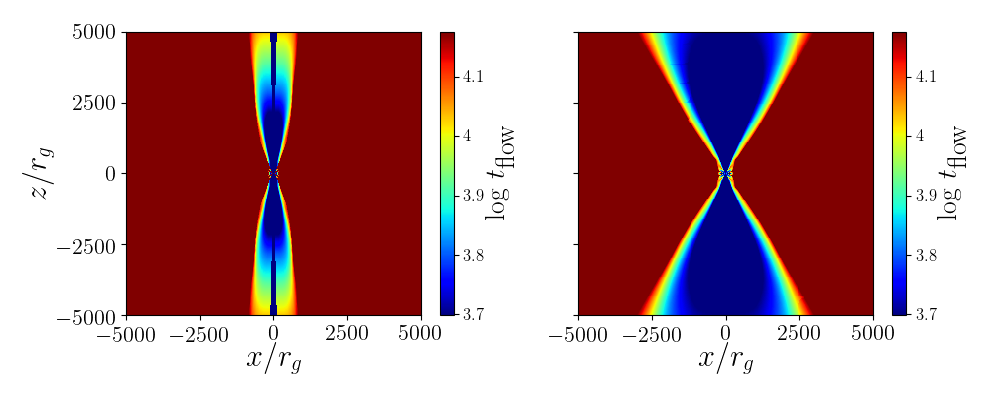}
    \caption{Flow time (colors) for models \texttt{s00} (left) and \texttt{m09} (right). The top panels are zoomed in to show the accretion flow. The white circles are at $r=20\,r_g$ and $r=25\,r_g$ for \texttt{s00} and $r=30\,r_g$ and $r=40\,r_g$ for \texttt{m09}. The bottom panels show the large scale flow time of the jet, wind, and disk. The boundary between the disk and wind regions roughly coincides with the edge of the converged region of the simulation so the wind and jet are evidently converged. The radial extent of the converged region of the disk taken near the equatorial plane is $\lesssim 40\,r_g$ using the flow time described in the text.}
    \label{fig:fig1}
\end{figure*}

\begin{figure}
	\includegraphics[width=\columnwidth]{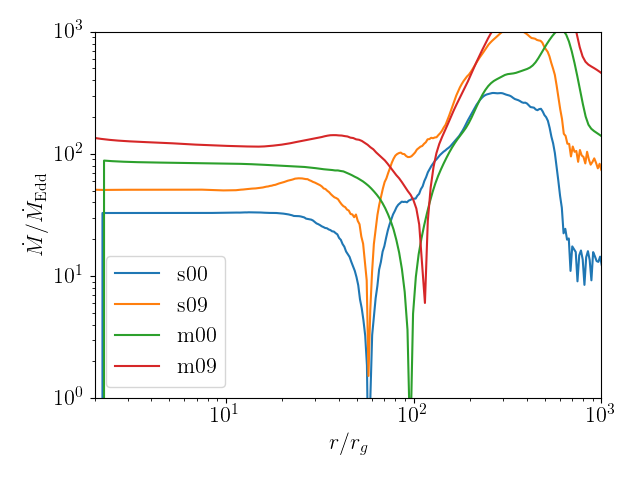}
    \caption{Time- and azimuth-averaged mass accretion rates as a function of radius for the four simulations.}
    \label{fig:fig2}
\end{figure}

When analyzing the simulation output from \textsc{KORAL}, we focus on the converged, steady-state regions of the flow.
As a diagnostic for convergence, we compute the flow time in each cell, 
\begin{equation}
t_{\rm{flow}} = \frac{r}{v_{\rm{pol}}},
\label{eq:tflow}
\end{equation}
where $r$ is the radius and $v_{\rm{pol}}=\sqrt{v_r^2+v_\theta^2}$ is the poloidal fluid velocity. Eliminating the first $5000\,t_g$ of the simulation, which is a transient phase during which the accretion rate is building up from zero, we consider the effective duration of each simulation to be 
\begin{equation}
t_{\rm{sim}} = t_{\rm{max}} - 5000\,t_g,
\label{eq:sim}
\end{equation}
where the $t_{\rm max}$ values are given in Table~\ref{tab:tab1}. We consider cells with $t_{\rm{flow}} < t_{\rm{sim}}$ to have reached a steady state.

Given the relatively short duration of the simulations discussed here, most regions of the disk have not reached steady state. However, most regions of the wind and jet are in a steady state. This can be seen in Figure \ref{fig:fig1} where we show $t_{\rm flow}$ (colors) for the simulations \texttt{s00} and \texttt{m09}. Deep blue regions correspond to $t_{\rm flow}<5000t_g$ and are safely in inflow equilibrium. Deep red regions correspond to $t_{\rm flow}>t_{\rm sim}$, and are certainly not in inflow equilibrium. Intermediate color regions have achieved partial inflow equilibrium, but for the purposes of this paper we consider them to be in steady state.

The two panels showing regions close to the BH ($r<50r_g$) indicate that the disk has reached inflow equilibrium out to a radius $\approx 20-25$ in \texttt{s00} and $\approx 30-40$ in \texttt{m09} (indicated by the white circles). The MAD simulation is in steady state out to a larger radius because of its larger magnetic field which gives stronger angular momentum transfer and hence a larger radial infall velocity in the disk. For the same reason, as Table~\ref{tab:tab1} and Figure~\ref{fig:fig2} show, the two MAD simulations have larger mass accretion rates compared to their SANE counterparts, even though both are initialized with the same torus. 

The two panels that show the large scale structure of the simulations ($r<5000r_g$) indicate that the region near the poles has reached steady state out to quite large radii. This region corresponds to the jet and the wind. Therefore we can study the properties of these two regions out to $r=5000r_g$. This is beyond the photosphere of the torus, which is at $r\sim2500r_g$ (see Figure \ref{fig:fig6} later), and hence allows us to study observational signatures of the jet/wind even for off-axis observers. The torus itself is far from reaching a dynamical steady state, though it is in hydrostatic equilibrium by construction (initial state). However, regions close to the photosphere of the torus have achieved energy balance (viscous heating balanced by radiative cooling) to an optical depth of several. Hence we can study the radiation emerging from the photosphere when we carry out the radiative transfer calculations described later.

In Figure \ref{fig:fig2} we show the time- and azimuth-averaged mass accretion rates for the four models. We see that model \texttt{s00} has reached a quasi-steady constant value of $\dot{M}$ out to $r_{\rm{eq}} \sim 25 r_g$, and that \texttt{m09} has achieved steady state out to $r_{\rm eq} \sim 40 r_g$. These estimates agree with the discussion in the previous paragraphs. 

Figure \ref{fig:fig2} and Table \ref{tab:tab1} indicate that the spin 0.9 models have $\dot{M}/\dot{M}_{\rm Edd}$ values about 50\% larger than the equivalent spin 0 models. This is caused by a combination of two factors. First, note that $\dot{M}_{\rm Edd}$ is defined via the radiative efficiency $\eta$ of a thin accretion disk (eq.~\ref{eq:mdotEdd}). Since the value of $\eta$ for spin 0.9 is 2.7 times greater than that for spin 0 (0.156 vs 0.0572), this effect by itself implies that, for the same physical mass accretion rate $\dot{M}$ ($\rm g\,s^{-1}$), the spinning BH models would have larger $\dot{M}/\dot{M}_{\rm Edd}$ by a factor of 2.7. However, this factor is partly counteracted by the fact that the ISCO and horizon radii are both substantially smaller for a spinning BH ($2.32r_g, ~1.44r_g$, respectively, vs $6r_g, ~2r_g$, for a non-spinning BH). Since super-Eddington accretion flows lose a lot of mass via winds, the smaller radii mean that a smaller fraction (by almost a factor of 2) of the available gas crosses the horizon. The combination of the two effects results in a net enhancement of $\dot{M}/\dot{M}_{\rm Edd}$ by 50\%.

\subsubsection{Models without a Relativistic Jet: Non-Jetted TDEs}

\begin{figure}
	\includegraphics[width=\columnwidth]{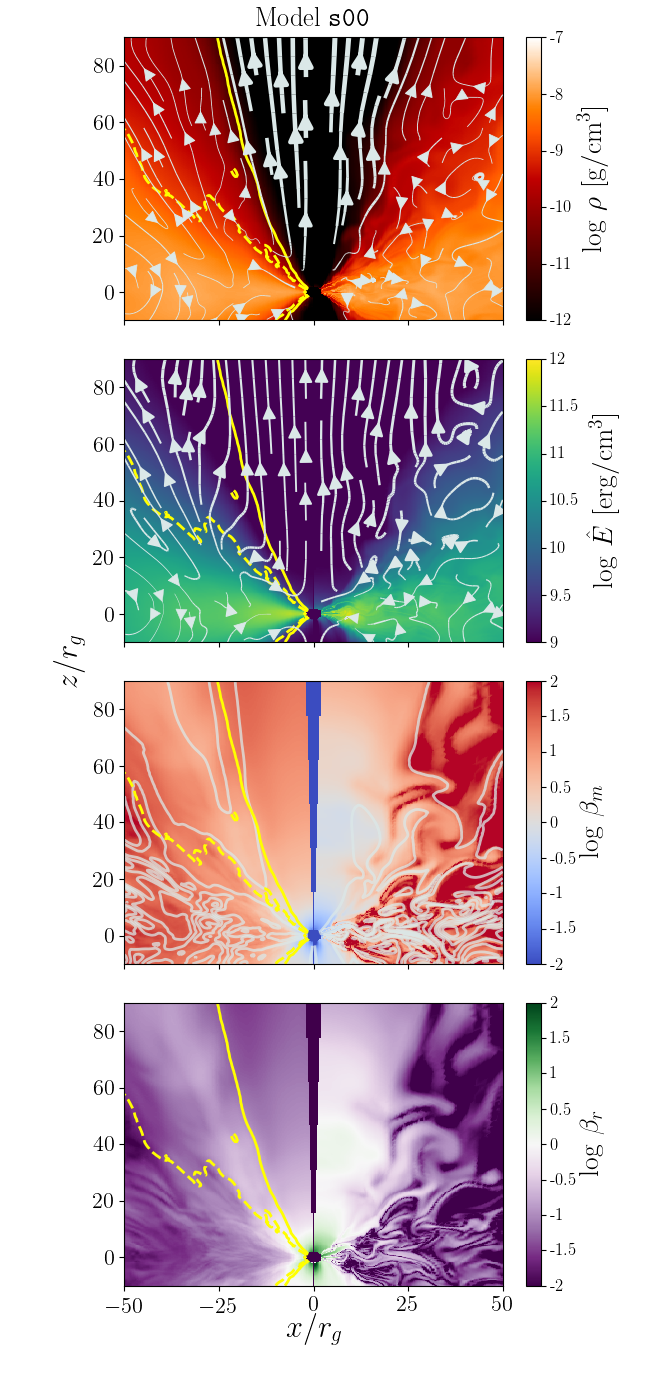}
    \caption{Fluid properties of the SANE model \texttt{s00}. The left half of each panel shows time-averaged properties ($t=15,000-20,000\,t_g$) and the right half shows properties of the snapshot at $t=20,000\,t_g$. Top panel: gas density (color scale) with fluid velocity (streamlines) superposed. Second panel: radiation energy density (color scale) with radiation flux (streamlines) superposed. Third panel: magnetic pressure ratio $\beta_m$ (color scale) with poloidal magnetic field lines (contours) superposed. Bottom panel: radiation pressure ratio $\beta_r$ (color scale). The yellow contours indicate the jet/wind boundary (Be=Be$_{\rm{crit}}$, solid yellow) and the wind/disk boundary (Be=$0$, dashed yellow).}
    \label{fig:fig3}
\end{figure}

\begin{figure}
	\includegraphics[width=\columnwidth]{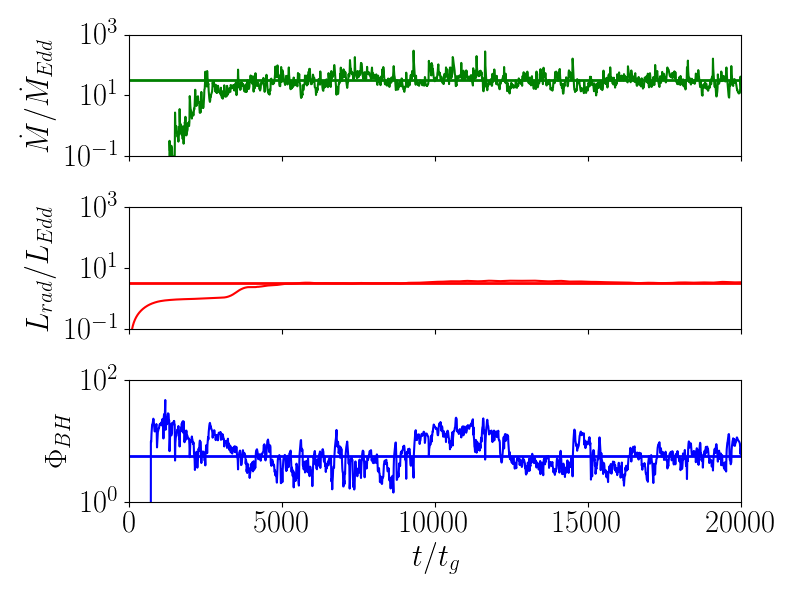}
    \caption{Mass accretion rate in units of $\dot{M}_{\rm Edd}$ (top), radiative luminosity in units of $L_{\rm Edd}$ (middle), and magnetic flux parameter $\Phi_{\rm{BH}}$ (bottom) as a function of time for the SANE accretion disk model \texttt{s00}. The solid lines mark quantities averaged over the last $5000\,t_g$ of the simulation which are $\dot{M}/\dot{M}_{\rm{Edd}}=32$ (top), $L_{\rm{rad}}/L_{\rm{Edd}}=3.4$ (middle), and $\Phi_{\rm{BH}}=6$ (bottom). The disk is evidently SANE for the entire simulation.}
    \label{fig:fig4}
\end{figure}

\begin{figure*}
	\includegraphics[width=\textwidth]{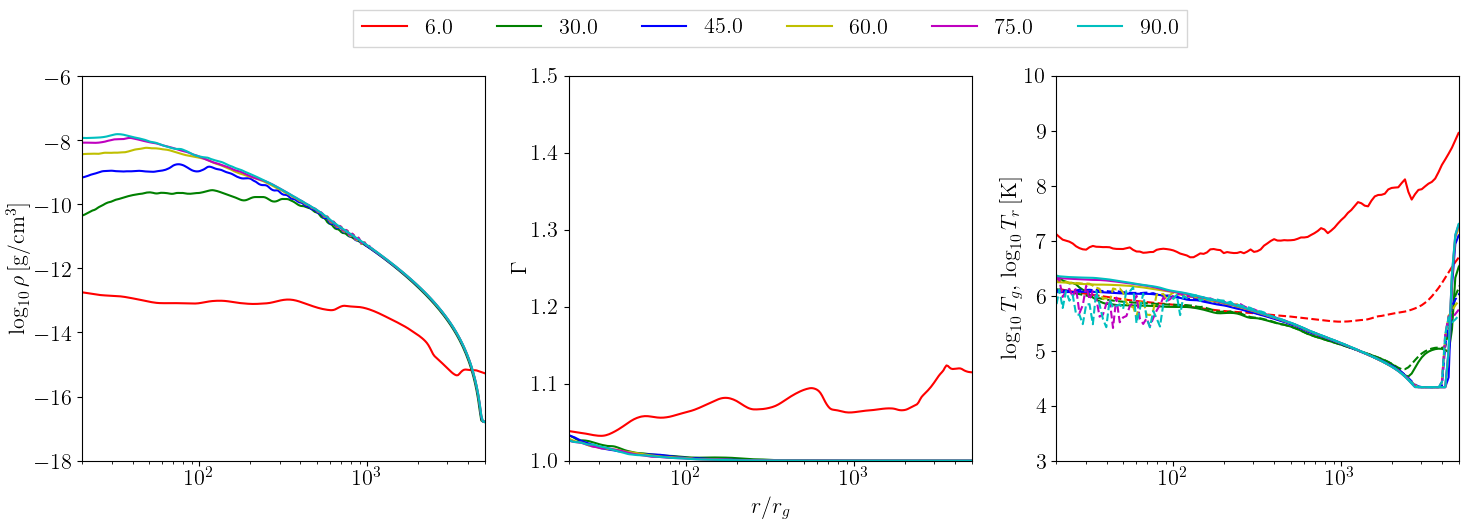}
    \caption{Radial profiles at various polar angles $\theta$ indicated by color (see legend at top) of gas density $\rho$ (left panel), Lorentz factor $\Gamma$ (middle panel), and gas (solid lines) and radiation (dashed lines) temperatures (right panel) for the SANE accretion disk model \texttt{s00}. Note that the gas and radiation temperature are obtained after post-processing with \textsc{HEROIC}.}
    \label{fig:fig5}
\end{figure*}

\begin{figure*}
	\includegraphics[width=\textwidth]{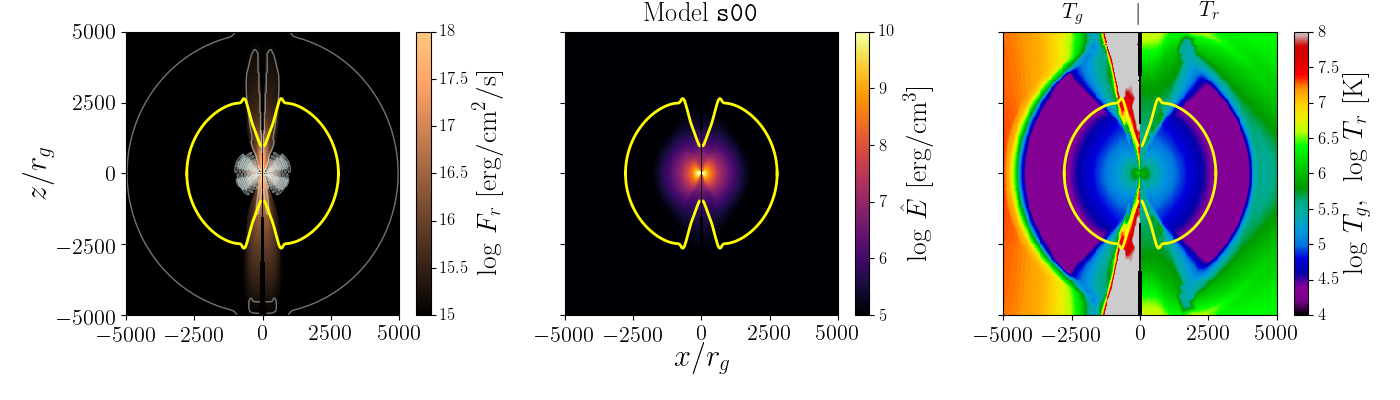}
	\includegraphics[width=\textwidth]{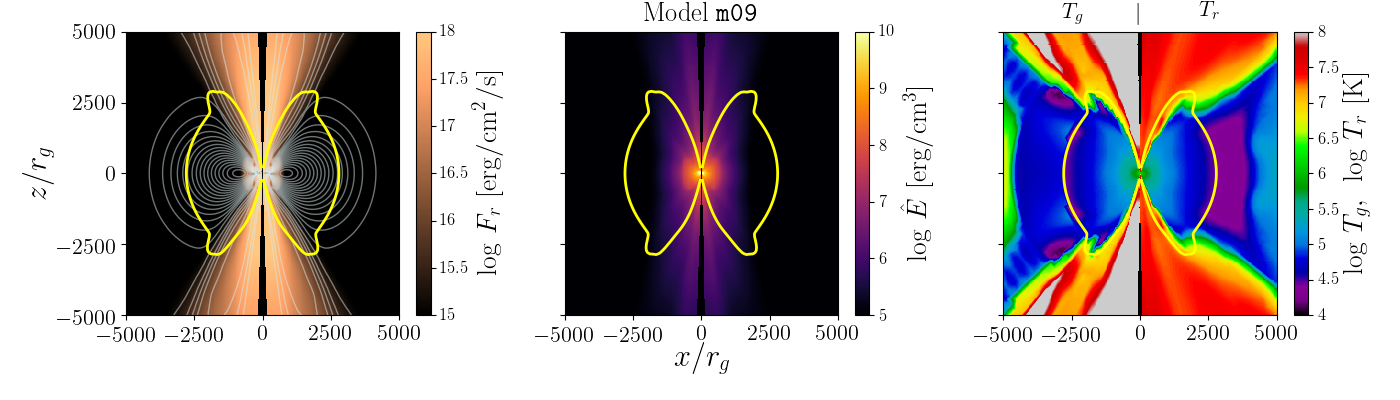}
    \caption{Large scale characteristics of models \texttt{s00} (top) and \texttt{m09} (bottom). In each panel, the yellow contour shows the electron scattering photosphere. Left panels: Show the radial flux of radiation (color scale) and vector potential ($A_\phi$, white contours). Model \texttt{s00} produces only a small amount of radiation in the jet, whereas \texttt{m09} launches a powerful beam of radiation. In the latter model, the lower velocity disk wind also carries a substantial amount of radiative energy. Middle panels: Show the radiation energy density. In model \texttt{s00}, much of the energy density is contained in the accretion flow and little is carried in the outflow. In model \texttt{m09}, the outflow region is strongly radiation dominated. Right panels: Show the gas (left) and radiation (right) temperature as obtained after post-processing with \textsc{HEROIC}. In both models, off-axis observers would see a thermal disk component with a temperature of $T_r\sim 10^{4.4}$ K, along with higher energy radiation from the wind and jet. On-axis observers would see some radiation from the disk and wind and a strong component of higher energy photons from the base of the jet, the funnel walls ($T\gtrsim 10^{5}$) K and the optically thin jet.}
    \label{fig:fig6}
\end{figure*}

In Figure \ref{fig:fig3} we show some properties of the simulation \texttt{s00}, which we consider to be our fiducial non-jetted TDE model. Models \texttt{s09} and \texttt{m00} also have similar properties and are equally valid models of non-jetted TDEs. Results corresponding to these other models are shown in Appendix \ref{sec:appC}. The right and left halves of the panels in Figure \ref{fig:fig3} show, respectively, a snapshot of simulation \texttt{s00} at $t=20,000t_g$, and time-averaged properties over $t=15,000-20,000t_g$.

The topmost panel shows the distribution of gas density (colors) and velocity (streamlines). The disk is evidently thick and turbulent. The gas accretes onto the black hole primarily along the equatorial plane. The flow shows significant turbulence even at relatively large radii. The turbulent structure of the velocity streamlines is the result of material near the black hole gaining energy and being launched back into the disk. For such a low binding energy disk, small perturbations can lead to the material becoming unbound quite easily \citep{Coughlin2014}. Outflows driven predominantly by radiation pressure are evident within $\sim45^\circ$ from the pole.

The second panel shows the radiation energy density (colors) and radiative flux (streamlines). Much of the radiation energy density is contained within the disk near the equatorial plane. Radiation is advected in with the accretion flow and escapes out through the funnel, driving a mildly relativistic outflow ($\Gamma \sim 1.1$). We will refer to the highest velocity material here as the `jet' even though it is too slow to be considered a true relativistic jet.

The third panel shows $\beta_m$, the ratio between gas plus radiation pressure and magnetic pressure (colors). The contours follow the vector potential ($A_\phi$), which maps the poloidal magnetic field structure. The torus is predominantly gas/radiation pressure dominated throughout, but there are pockets of magnetic field dominance in the jet region.

In the bottom panel, we show $\beta_r$, the ratio between gas plus magnetic pressure and radiation pressure. The disk and wind are evidently radiation dominated while the jet has nearly equal magnetic and radiation pressures.

In each panel in Figure \ref{fig:fig3}, we show the contour corresponding to ${\rm Be} = {\rm{Be}}_{\rm{crit}}$ as the solid yellow line, and ${\rm Be}=0$ as the dashed yellow line. We find that our choice of ${\rm{Be}}_{\rm{crit}}=0.05$ divides the simulation appropriately between the faster jet and slower wind. This contour also divides the simulation into the magnetic pressure dominated jet and radiation/gas pressure dominated wind (and disk). We estimate the jet opening angle $\theta_j$ using the contour of ${\rm{Be}}_{\rm{crit}}$ at large radii ($r > 1000\,r_g$). For models \texttt{s00} and \texttt{s09}, we find a similar jet opening angle of $\theta_j \sim 12^\circ$. For model \texttt{m00}, we find that the outflow actually becomes quite optically thick. Furthermore, the simulation appears to have only launched a wind, with hardly any  `jet' (see below).

In Figure \ref{fig:fig4} we show for model \texttt{s00} the mass accretion rate as a function of time in units of $\dot{M}_{\rm{Edd}}$ (top), the radiative luminosity $L_{\rm{rad}}$ in units of $L_{\rm{Edd}}$ (middle), and the magnetic flux parameter $\Phi_{\rm BH}$ as given by equation (\ref{eq:phiBH}). The average accretion rate over the last $5000\, t_g$ of the simulation is roughly $\sim 32$ times the Eddington rate. In contrast, the radiative luminosity [equation (\ref{eq:Lrad})], computed outside the photosphere ($r \sim 2500\, r_g$, see Figure \ref{fig:fig6} below) is roughly $3.4 L_{\rm{Edd}}$. Thus the accretion flow is radiatively very inefficient, consistent with expectations for this regime of accretion (see the discussion of the slim disk model in \citealt{Abramowicz1988}). As for the magnetic flux parameter, the initial setup of multiple quadrupolar field  loops prevents the accumulation of much magnetic field (of single polarity) around the BH. This leads to a low magnetic flux parameter $\Phi_{\rm BH}\sim6$ even at late times, which is much smaller than that expected for a MAD system ($\Phi_{\rm BH} \sim 40-50$). Hence model \texttt{s00} belongs firmly to the class of SANE accretion flows.

In Figure \ref{fig:fig5} we show radial profiles of density, Lorentz factor, gas temperature and radiation temperature, corresponding to various polar angles $\theta$. The gas density in the jet ($\theta=6^\circ$) is roughly four orders of magnitude less than that in the disk. The jet is optically thin and the gas here can be accelerated by radiation leaking towards the pole from the funnel wall. However, despite the low gas density, the Lorentz factor only goes up to $\Gamma\sim 1.1$ at large radii. The velocity is much smaller in the wind.

The third panel in Figure \ref{fig:fig5} and the right panel in Figure \ref{fig:fig6} indicate that the gas and radiation temperature (as obtained after post-processing with \textsc{HEROIC}) track each other closely in the disk and wind but deviate significantly in the jet. In addition, the radiation temperature at the photosphere (see Figure \ref{fig:fig6}) is noticeably hotter in the wind and jet in comparison to the torus. This is the case for three reasons: (i) viscous heating is largest in the jet, wind, and funnel walls and the gas temperature in these regions is elevated compared to the torus, where there is negligible heating, (ii) the wind is optically thick and the gas and radiation have come into equilibrium, (iii) the jet is optically thin, so the radiation and gas remain out of equilibrium; however, Compton scattering of soft photons still elevates the radiation temperature here significantly. Radiation in the jet is dominated by flux coming out of the cooler funnel wall. Given the low optical depth of the jet, this radiation is only mildly Comptonized by the hotter jet gas, hence $T_r \ll T_g$. If we focus on the outer photosphere of the torus ($r\sim2500\,r_g$), the emerging radiation at $\theta = 45^\circ$ and higher has a temperature $T_r\sim10^{4.4}$\,K, the radiation in the wind and jet ($\theta\lesssim 30^\circ$) has a temperature of $10^{5-6}$\,K. As we discuss later, there are signatures of emission from all of these regions in the spectra we calculate from this model.

The three upper panels in Figure \ref{fig:fig6} show the large scale properties of model \texttt{s00}. The location of the electron scattering photosphere is indicated by the yellow contours. The radiation temperature in the last panel confirms the discussion in the previous paragraph. For a wide range of angles surrounding the equator, $T_r\sim 10^{4.4}$ K (purple color), so the escaping radiation here will be in the optical/UV band. For a range of angles closer to the pole, the temperature is around $10^5$\,K (light blue). This region corresponds to the slow-moving wind. Finally, close to the pole, the temperature goes up to $\sim10^6$\,K. This is the jet. How much radiation each of these three zones contributes to the observed spectrum depends on the gas density and temperature (at the photosphere) and also the viewing angle, as we will discuss later. 

\begin{figure}
	\includegraphics[width=\columnwidth]{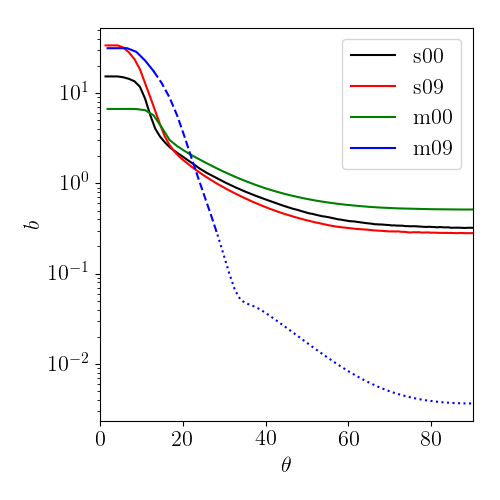}
    \caption{Beaming factor as a function of $\theta$ for the four models, computed at $r=3000r_g$ (outside of the photosphere). For model \texttt{m09}, we break the curve into the disk/wind (dotted line), jet sheath (dashed line), and jet core (solid line).}
    \label{fig:fig7}
\end{figure}

\begin{figure*}
	\includegraphics[width=\textwidth]{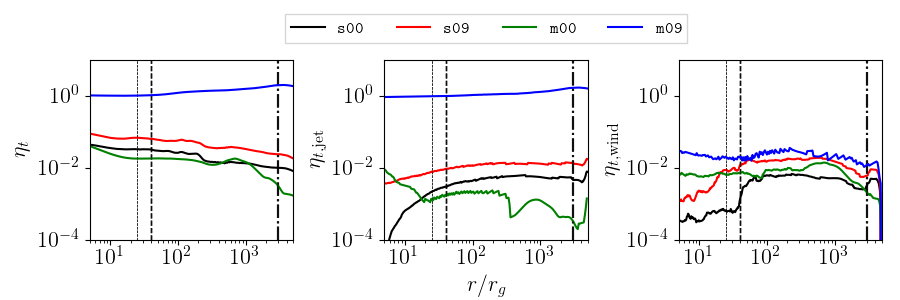}
    \caption{Radial profiles of efficiencies for the four models. Left panel: Total efficiency $\eta_t$ from all sources of energy. Middle panel: Efficiency $\eta_{t,\rm{jet}}$ corresponding to the energy carried by the jet. Right panel: Efficiency $\eta_{t,\rm{wind}}$ corresponding to the energy carried by the wind. We compute $\eta_t$ at $r=25\,r_g$ (thin dashed line) for SANE models (\texttt{s00, s09}) and $r=40\,r_g$ (dashed line) for MAD models (\texttt{m00, m09}). The jet and wind are in steady state to much larger radii (see Fig.~\ref{fig:fig1}) so we compute $\eta_{t,\rm{wind}}$ and $\eta_{t,\rm{jet}}$ outside of the photosphere ($r=3000\,r_g$, dash-dotted line). Most of the energy in the non-jetted models, \texttt{s00, s09, m00}, is contained within the disk. The jet and wind efficiencies of these models are much smaller than the total efficiency. The jetted model, \texttt{m09}, has a substantial total efficiency $\eta_t>100$\%. Moreover, $\eta_t\approx\eta_{t,\rm{jet}}$, which means that most of the luminosity comes out in the jet.}
    \label{fig:fig8}
\end{figure*}

We find it useful to define the beaming factor $b$, which is the ratio of the isotropic equivalent luminosity along a given direction $\theta$ to the total radiation luminosity, $b=L_{\rm{iso}}/L_{\rm{rad}}$. We compute this quantity at $r=3000\,r_g$ (somewhat outside the photosphere) and show the results in Figure \ref{fig:fig7}. Leaving aside model \texttt{m09}, which we discuss later, we see that the other three models, \texttt{s00}, \texttt{s09} and \texttt{m00}, all have similar behavior. The beaming factor is largest at the poles and drops by an order of magnitude at the equator. Most of the radiation escapes within $\sim 15^\circ$, which agrees fairly well with our definition of the jet boundary. The steady decline of $b$ between $\theta\sim15^\circ$ and $\sim40^\circ$ is due to the wind, which allows some radiation to escape. Beyond this angle, we have the torus, and here the beaming is effectively independent of angle.

Finally we discuss various efficiencies. The radiative efficiency $\eta_r = L_{\rm rad}/\dot{M}c^2$, computed just outside the photosphere, is $\sim 0.6$\% for model \texttt{s00}, $\sim 1.2$\% for model \texttt{s09}, and $\sim 0.1$\% for model \texttt{m00}. All three models are clearly radiatively inefficient, with efficiencies a tenth or less of the standard Novikov-Thorne efficiency of a thin accretion disk (5.7\% for $a_*=0$ and 15.6\% for $a_*=0.9$).

The total efficiency $\eta_t$, which is computed from the total luminosity including all forms of energy, is a different story. We show this quantity in Figure \ref{fig:fig8} for the entire flow (left panel) as well as for the jet $\eta_{t,\rm{jet}}$ (middle panel) and wind $\eta_{t,\rm{wind}}$ (right panel) separately. Given that the wind/jet are converged out to much larger radii than the disk, for each simulation we compute a representative $\eta_t$ at the extent of the converged region (see Figure \ref{fig:fig1}) while $\eta_{t,\rm{jet}}$ and $\eta_{t,\rm{wind}}$ are computed outside of the photosphere at $r=3000\,r_g$. Thus, for the purposes of this discussion, $\eta_t$ is a useful diagnostic for how efficiently rest mass energy is converted to other forms of energy while the ratio $\eta_{t,\rm{jet}}/\eta_{t,\rm{wind}}$ describes how much energy escapes in the jet versus the wind. For model \texttt{s00} we find $\eta_t \sim 3$\%, for \texttt{s09} $\eta_t \sim 7\%$, and for model \texttt{m00} $\eta_t \sim 2$\%. By this measure all three flows are reasonably efficient, with total efficiencies of around half the standard thin disk efficiency. The radiative efficiency in each model is nearly 10 times smaller than the total efficiency. This is because, in super-Eddington accretion flows, the outflowing energy is primarily in the form of gravitational, binding, and magnetic energy (see \citealt{Sadowski2016c}). This is particularly true of the disk, where radiation is trapped because of the large optical depth and photons are advected in with the accreting gas. Considering the efficiencies in the jet and wind, we find $\eta_{t,\rm{jet}}/\eta_{t,\rm{wind}} \approx 1.26$ for model \texttt{s00} and $\eta_{t,\rm{jet}}/\eta_{t,\rm{wind}} \approx 2.29$ for model \texttt{s09}. These two models have nearly equal amounts of energy escaping in the jet and the wind. By contrast, model \texttt{m00} is evidently a wind dominated system, since $\eta_{t,\rm{jet}}/\eta_{t,\rm{wind}} \approx 0.17$.

The various efficiencies discussed here are listed in Table~\ref{tab:tab1}.

\subsubsection{Model with a Relativistic Jet: Jetted TDE}
\label{sec:sec442}


\begin{figure}
	\includegraphics[width=\columnwidth]{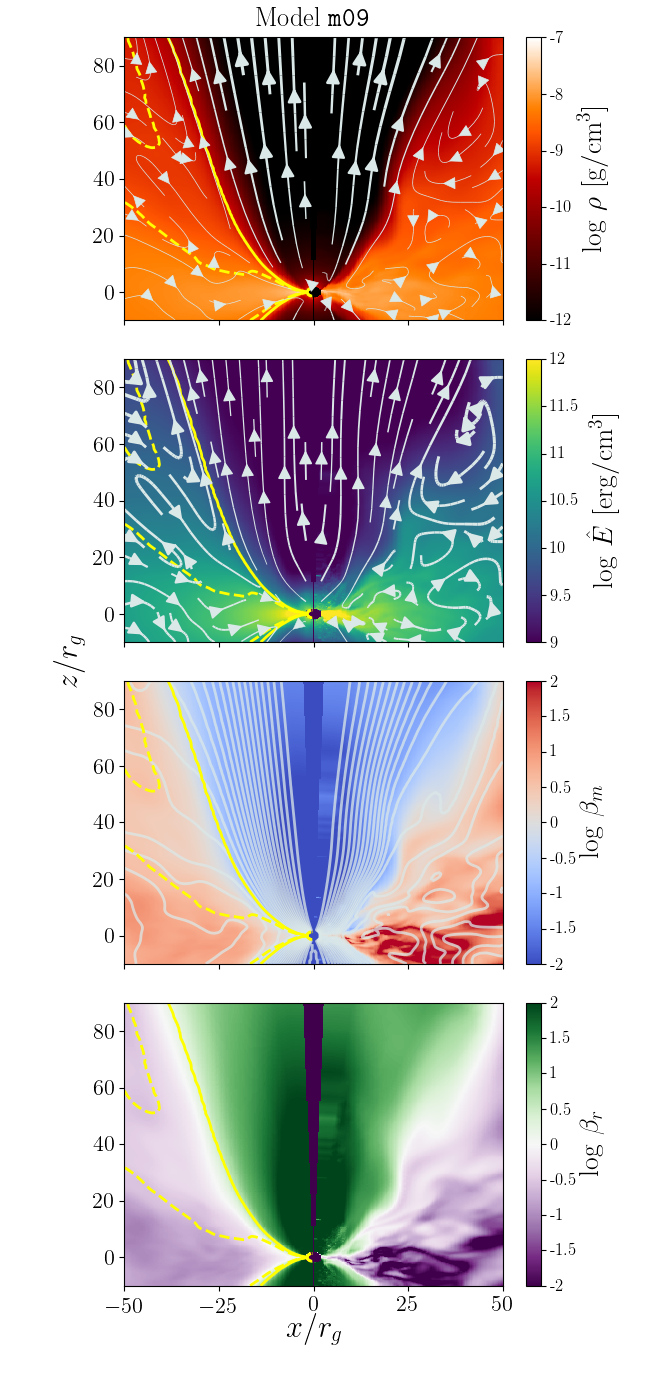}
    \caption{Fluid properties of the MAD model \texttt{m09}. The left half of each panel shows time-averaged properties ($t=20,000-25,000\,t_g$) and the right half shows properties of the snapshot at $t=25,000\,t_g$. Top panel: gas density (color scale) with fluid velocity (streamlines) superposed. Second panel: radiation energy density (color scale) with radiation flux (streamlines) superposed. Third panel: magnetic pressure ratio $\beta_m$ (color scale) with poloidal magnetic field lines (contours) superposed. Bottom panel: radiation pressure ratio $\beta_r$ (color scale). The yellow contours indicate the jet/wind boundary (Be=Be$_{\rm{crit}}$, solid yellow) and the wind/disk boundary (Be=$0$, dashed yellow).}
    \label{fig:fig9}
\end{figure}

\begin{figure}
	\includegraphics[width=\columnwidth]{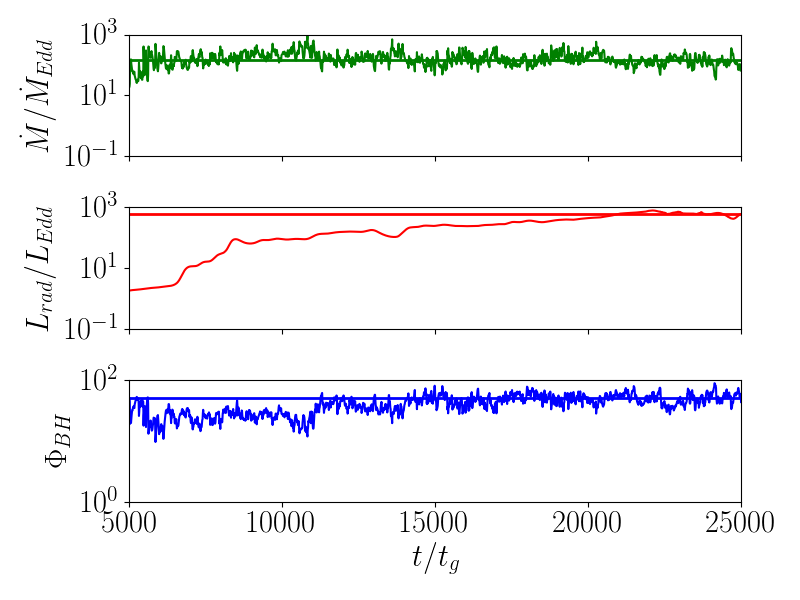}
    \caption{Mass accretion rate in units of $\dot{M}_{\rm Edd}$ (top), radiative luminosity in units of $L_{\rm Edd}$ (middle), and magnetic flux parameter $\Phi_{\rm{BH}}$ (bottom) as a function of time for the MAD accretion disk model \texttt{m09}. The solid lines mark quantities averaged over the last $5000\,t_g$ of the simulation which are $\dot{M}/\dot{M}_{\rm{Edd}}=150$ (top), $L_{\rm{rad}}/L_{\rm{Edd}}=615$ (middle), and $\Phi_{\rm{BH}}=51$ (bottom). The disk is evidently in the MAD state.}
    \label{fig:fig10}
\end{figure}

\begin{figure*}
	\includegraphics[width=\textwidth]{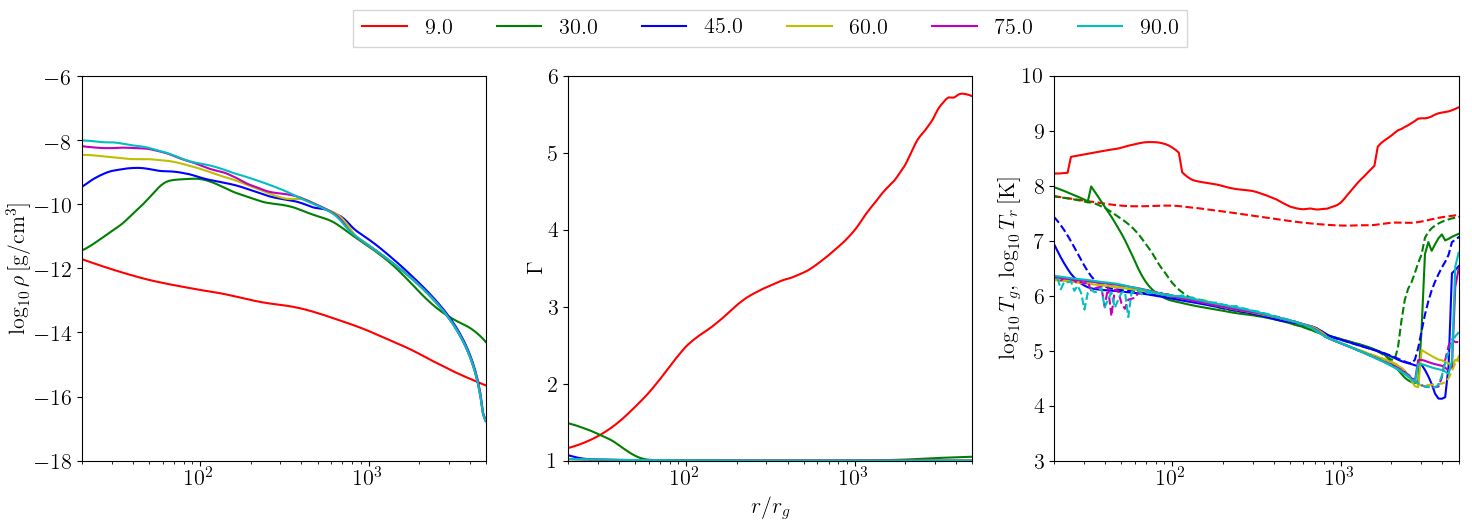}
    \caption{Radial profiles at various polar angles $\theta$ indicated by color (see legend at top) of gas density $\rho$ (left panel), Lorentz factor $\Gamma$ (middle panel), and gas (solid lines) and radiation (dashed lines) temperatures (right panel) for the jetted TDE model \texttt{m09}.}
    \label{fig:fig11}
\end{figure*}


The simulation \texttt{m09} behaves very differently from the other three models. Notably, it shows strong jet-related features. For this reason, we consider it our fiducial model of a jetted TDE. In Figure \ref{fig:fig9} we show some properties of this model. The left half of each panel shows the time and azimuth averaged properties taken over $t=20,000-25,000\,t_g$, while the right half shows the state of the simulation at $t=25,000\,t_g$.

The four panels correspond to the same quantities as in the case of the non-jetted model \texttt{s00} discussed earlier (Figure \ref{fig:fig3}). The gas density (top panel) and radiation energy density (second panel) are both slightly lower in the case of \texttt{m09}, despite the larger $\dot{M}$. The magnetic pressure ratio $\beta_m$ (third panel) clearly shows that the funnel region is dominated by magnetic pressure. This is natural since this model is in the MAD state and is much more magnetized than model \texttt{s00}. Similarly, the bottom panel shows that radiation pressure is negligible in the funnel compared to the other pressures (notably magnetic). 

Figure \ref{fig:fig10} shows the same quantities as Figure \ref{fig:fig4}, but now for model \texttt{m09}. The average accretion rate over the last $5000\, t_g$ of the simulation is roughly $150$ times the Eddington rate. This model is initialized with a single large scale dipolar loop, which causes significant accumulation of magnetic field around the BH with a single polarity. The accretion flow thus becomes MAD already around $t=5000\, t_g$, and remains MAD throughout the rest of the simulation ($\Phi_{\rm BH} \sim 50$). 

The radiative luminosity as measured at $r \sim 3000\, r_g$ is shown in the middle panel of Figure \ref{fig:fig10}. The luminosity rapidly rises to $\sim 100 L_{\rm{Edd}}$ early on, this rise coinciding with the ultra-relativistic jet head crossing the radius where we compute the luminosity. The luminosity then continues to increase slowly until it finally saturates at $\sim 600 L_{\rm{Edd}}$ at $t = 25,000 \, t_g$.

Note the huge difference in radiative luminosity between the jetted TDE model \texttt{m09} we are discussing here and the non-jetted TDE models \texttt{s00, s09, m00} discussed in the previous subsection. The non-jetted models all had luminosities of only a few Eddington, despite having highly super-Eddington mass accretion rates. That is consistent with theoretical expectations for super-Eddington accretion, e.g., the slim disk model \citep{Abramowicz1988}, and implies very radiatively inefficient accretion. In contrast, not only is the jetted TDE model \texttt{m09} radiatively efficient, it is in fact super-efficient in the sense that an accretion rate of $150\dot{M}_{\rm Edd}$ gives a luminosity of not just $150L_{\rm Edd}$, but $600L_{\rm Edd}$; the radiative efficiency is thus a factor of 4 larger than the efficiency of a thin accretion disk around a BH of the same spin.

Figure \ref{fig:fig11} is similar to Figure \ref{fig:fig5}, but now refers to model \texttt{m09}. The central region of the jet ($\theta=9^\circ$) is accelerated to a fairly large Lorentz factor $\Gamma \sim 5-6$, while at larger angles from the pole the Lorentz factor drops off rapidly. Motivated by the observation that the late-time radio emission from J1644 is likely due to a two component jet \citep{Berger2012,Wang2014,Mimica2015,Liu2015}, we define two jet regions. We call the higher Lorentz factor zone the jet `core' and the rest of the jet the `sheath', defining the boundary between the two to be located at $\Gamma = 2$ (following \citealt{Mimica2015}). With this definition, the boundary between the core and the sheath is at $\theta_{j,c} \sim 15^\circ$, while the boundary between the sheath and the wind (defined by ${\rm Be = Be_{crit}}$) is at $\theta_{j,s} \sim 30^\circ$.  

The lower panels in Figure \ref{fig:fig6} show the large scale properties of model \texttt{m09}. Outside of the jet and wind, the electron scattering photosphere is at $\sim 2500\, r_g$, similar to model \texttt{s00}. However, the jet and wind regions look noticeably different. For angles close to the axis, i.e., the jet region, the gas in \texttt{m09} is largely optically thin and the electron scattering photosphere penetrates all the way down to the base of the jet (r $\sim 25 r_g$). In contrast, the gas in \texttt{s00} becomes optically thick near $\sim 1000\,r_g$; however, we note that the photosphere depth of \texttt{s09} in the jet is at $\sim 50 r_g$ (see \ref{fig:figC7}), which is quite similar to that of \texttt{m09}. At larger angles, the photosphere in model \texttt{m09} extends to radii well outside the torus. This model ejects a lot of gas in the wind which forms an optically thick region surrounding the jet. There is no comparable feature in model \texttt{s00}. 

The jet in \texttt{m09} has a radiative flux that is roughly 2-3 orders of magnitude greater than the flux in \texttt{s00}. Also, the jet covers a much wider angle. Whereas \texttt{s00} showed hardly any radiation energy density outside of the optically thick regions, \texttt{m09} carries a significant amount of radiation energy density throughout the jet, signifying that photons escape more easily through the funnel and that synchrotron and inverse Compton processes are more effective at transferring energy from the gas to radiation. The radiation temperature in \texttt{m09} also shows significant differences. While the radiation properties of the torus are similar ($T_r \sim 10^{4.4}$\,K), the wind is characterized by a range of temperatures, $T_r \sim 10^{4.4}-10^7$\,K; the jet is slightly hotter than the hottest region of the wind, $T_r \sim 10^{7.3}$\,K. Edge-on observers are thus expected to see thermal emission in the optical/UV from the torus with a temperature of $T\sim 10^{4.4}$ K, and X-rays from the wind photosphere at temperatures up to $10^7$\,K. For face-on observers, since the funnel is optically thin to electron scattering, high energy photons from the entire length of the jet (down to $\sim 25 r_g$) will be seen, as well as Compton-upscattered photons in the hot gas. This emission will be strongly beamed because of the relativistic motion of the jet, and the jet will dominate the observed radiation. A face-on observer will also receive radiation from the wind and the torus, but there is little beaming so their contribution will be a sub-dominant component of the observed radiation. The emission properties are discussed in more detail later.

The much stronger beaming in model \texttt{m09} relative to the other three models is also evident in Figure~\ref{fig:fig7}. Note that the beaming factor in the jet sheath, jet core, and disk/wind region of the simulation are plotted as a solid, dashed, and dotted line for \texttt{m09}. The beaming factor drops by two orders of magnitude between the pole and the jet sheath boundary (marked by the extent of the dashed line) and more than three orders of magnitude between the poles and the equator. Thus, in this model, most of the radiation escapes within $\sim 30$ degrees of the axis, which coincides with our definition of the jet boundary. 

We discuss next various efficiencies in model \texttt{m09}. The radiative efficiency is $\eta_r \sim 64$\%, i.e., about 4 times the efficiency of an equivalent thin accretion disk, as already discussed. Figure \ref{fig:fig8} shows the other efficiencies. The total efficiency (all energy forms) is $\eta_t \approx 105$\%, which is only modestly larger than the radiative efficiency. In contrast to the non-jetted models, where most of the energy emerges in forms other than radiation, the energy output of the jetted TDE model is dominated by radiation.  In addition, the jet efficiency $\eta_{t,\rm{jet}}$ is nearly the same as the total efficiency ($\eta_{t,\rm{jet}}\sim 100$\% at $r=40\,r_g$). Thus, nearly all the luminosity emerges in the jet. 

All of these features are the result of the fact that, among the four models considered in this paper, model m09 uniquely combines both a spinning black hole and MAD-level magnetic field strength. This combination is especially conducive to energy extraction from black hole spin via the Blandford-Znajek mechanism. As a result, in model m09, the total efficiency is unusually large, and moreover almost all the energy comes out in the jet. Furthermore, much of the jet energy gets converted into gas thermal energy and comes out in the form of radiation via inverse-Compton scattering. The low optical depth of the funnel region enables this radiation to escape. The other three models have negligible jet luminosity, so the funnel gas in those models is heated less and radiation from the funnel is less important.

The wind efficiency in model m09 is negligibly small compared to the jet efficiency: $\eta_{t,\rm{jet}}/\eta_{t,\rm{wind}} \approx 136$. Note again the large difference between model \texttt{m09} and the three other non-jetted TDE models, which have $\eta_{t,\rm{jet}} \approx \eta_{t,\rm{wind}}$ (in two models) and $\eta_{t,\rm{jet}} \ll \eta_{t,\rm{wind}}$ (in the third model).

It is worth highlighting that, among the four simulations we have described in this paper, only one is jet-dominated. What is unique about this model is that it is {\it both} MAD {\it and} has a rapidly spinning BH. Either of these features alone is not enough, e.g., the MAD model \texttt{m00} and the rapidly spinning BH model \texttt{s09} are essentially non-jetted. Only when both features are present, as in \texttt{m09}, does a powerful jet emerge.

Another notable point is that the total efficiency of model \texttt{m09} is 105\% or even slightly larger. This is reminiscent of the 140\% efficiency obtained by \citet{Tchekhovskoy2011} in their GRMHD simulation of a MAD rapidly spinning BH. Such efficiencies cannot be generated purely by accretion. Some of the luminosity must be emerging directly from the BH spin energy. There is thus a strong case for the Blandford-Znajek mechanism, or something akin to it, operating in model \texttt{m09}.

The many unique features of the jetted TDE model \texttt{m09} are consistent with the suggestion of \citet{Tchekhovskoy2014} that jetted TDEs such as J1644 must be MAD systems with rapidly spinning BHs. Those authors proposed their model on the basis of non-radiative GRMHD simulations (specifically \citealt{Tchekhovskoy2011}). Here we show via a GRRMHD simulation with full treatment of radiation that their proposal is indeed correct.

\section{Spectra and Comparisons with Observations}

\subsection{Radiation post-processing}
We post-process the \textsc{KORAL} simulation data using the radiative transfer code \textsc{HEROIC} described in \S\ref{sec:HEROIC}, and compute model spectra of the simulated systems. For all models, we use simulation data time-averaged over the final $5000t_g$. In the case of the 3D models, \texttt{m00} and \texttt{m09}, we also azimuthally-average the data. All models show quasi-steady behavior during the selected time interval. Also, their jets and winds have emerged outside the photosphere, so these components are able to contribute to the radiation seen by distant observers.

\subsection{Model Spectra}
\begin{figure}
	\includegraphics[height=0.9\textheight]{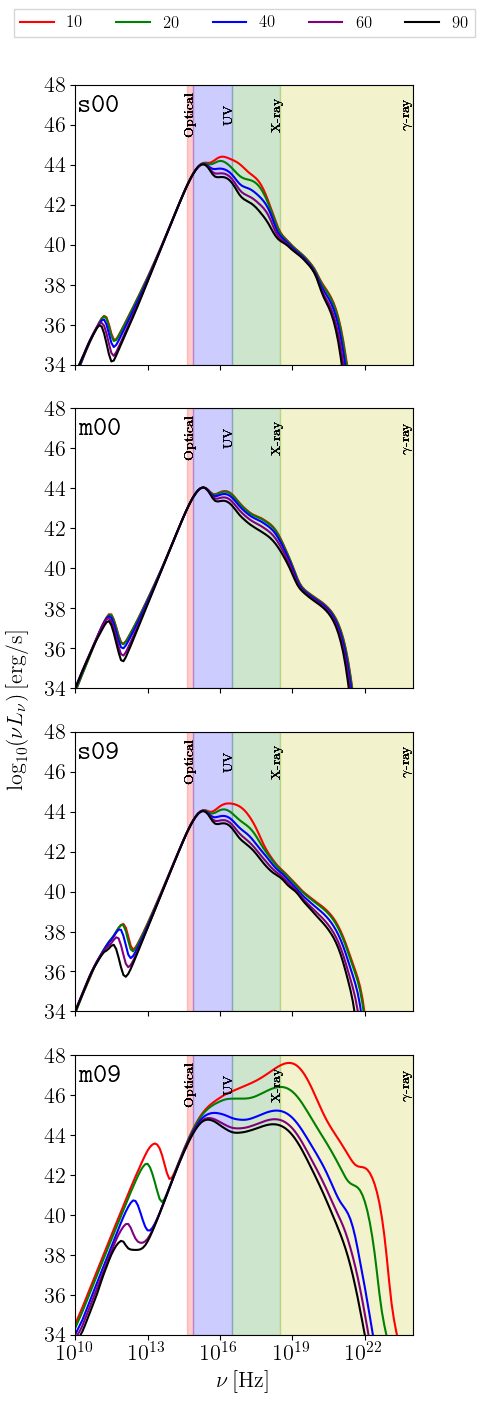}
    \caption{Spectra of the four models for observers at different viewing angles $\theta$ (indicated by color, see legend at top). The vertical colored bands indicate different regions of the spectrum.}
    \label{fig:fig12}
\end{figure}

In Figure \ref{fig:fig12} we show the spectra of the four models as seen by distant observers at different viewing angles. Spectra computed for the two non-jetted SANE models \texttt{s00} and \texttt{s09} are qualitatively quite similar, and can be decomposed into four peaks, one in the infrared (IR), one in near-UV, one in far-UV/soft X-rays and one in hard X-rays. The first peak (near-UV) is from the $T_r\sim 10^{4.4}$\,K photosphere of the torus, the second (far-UV) is from hotter gas ($T_r\sim 10^{5.5}$\,K)in the funnel wall/wind, and the third peak (hard X-rays) is from the hottest gas, which is in the jet. A tail of very high energy gamma-ray emission comes from Compton-upscattered radiation by hot gas, but this component is extremely weak. Thermal synchrotron emission from the jet produces the low luminosity ($< 10^{38}$ $\rm{erg\ \, s^{-1}}$) peak in the IR.

In models \texttt{s00}  and \texttt{s09}, the near-UV peak is essentially independent of viewing angle, as expected for a quasi-spherical optically-thick photosphere. This component has approximately an Eddington luminosity ($L_{\rm Edd} \approx 10^{44}\,{\rm erg\,s^{-1}}$ for a $10^6M_\odot$ BH), as we would expect for a radiation-dominated system. The other two components show viewing angle dependence in the case of models \texttt{s00, s09}. Observers looking down the radiation driven outflow ($\theta=10^\circ, ~20^\circ$) see strong emission from both the wind and the jet, the latter mildly enhanced by relativistic boosting. The spectra at these angles peak in far-UV/soft-X-rays, showing that they are dominated by the wind/jet. With increasing viewing angle, the contribution of the wind emission declines, and that of the jet emission declines even more strongly. At higher angles of inclination, the jet and the funnel wall become obscured by the torus material and the observed radiation in soft and hard X-rays is primarily from the part of the wind that pokes out beyond the torus photosphere.  The bolometric luminosity for a $90^\circ$ viewing angle is $\approx1.5\,L_{\rm{Edd}}$, most of it coming from the torus.

\begin{figure*}
  \centering
  \includegraphics[width=0.31\textwidth]{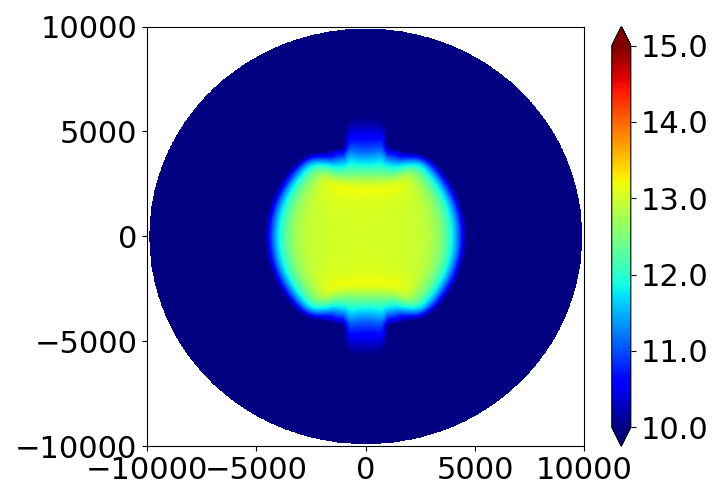}
  \includegraphics[width=0.31\textwidth]{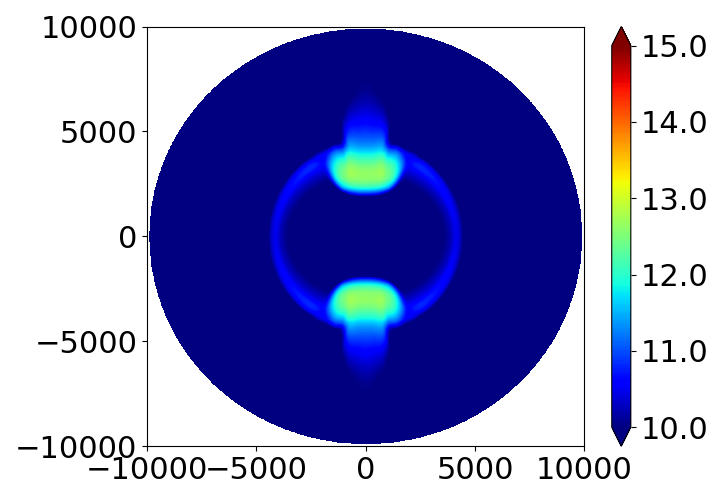}
  \includegraphics[width=0.31\textwidth]{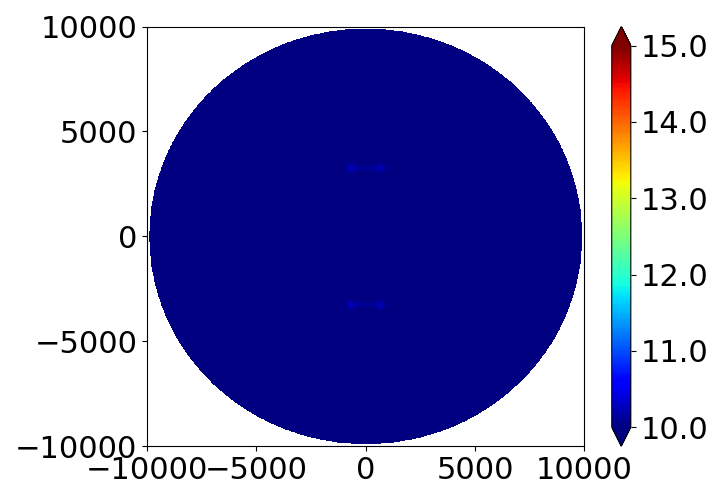}
  \caption{Images of model \texttt{s00} when viewed edge-on ($\theta=90^\circ$) in the optical/UV band (0.002-0.009 keV, left), X-ray band (0.2-10 keV, middle), and $\gamma$-ray band (15-150 keV, right).}
    \label{fig:fig13}
\end{figure*}

In Figure \ref{fig:fig13} we illustrate the above points via images of the fiducial model \texttt{s00} for an observer at viewing angle $90^\circ$. The optical/UV emission (left panel)  is clearly dominated by the large scale torus with a small contribution from the outflow. In the X-ray band (middle panel), the torus contributes virtually nothing, and the inner regions of the jet and wind (radii below $2500r_g$) are obscured by the torus. The dominant source of X-rays is emission from the base of the wind where it emerges outside the torus photosphere. In the $\gamma$-ray band (right panel), there is virtually no emission since none of the visible regions of the system are hot enough. Model \texttt{s09} is similar, but with slightly more X-ray emission in the outflow. The key point of this discussion is that non-jetted TDEs produce X-rays that are visible in all directions, but this radiation does not come from near the BH or from a corona surrounding the inner disk. Rather, the X-rays are from hot material in the outflow and come from regions that are thousands of $r_g$ from the BH. It is only out here that the X-rays are finally able to escape without being absorbed by the torus.

Our third non-jetted TDE model \texttt{m00} is similar in many respects to the two models discussed above, except that in this case even the funnel is optically thick (see the shape of the photosphere in Figure~\ref{fig:figC7}). As a result, the emission is nearly isotropic in all bands, and there is very little boost of the jet and wind emission for on-axis observers. 
Optical/UV radiation from the torus dominates at all angles, and the isotropic equivalent luminosity varies only a small amount with angle,
going from $2.5\,L_{\rm{Edd}}$ at the poles to $1.5\,L_{\rm{Edd}}$ at the equator. 

The model in \citet{Dai2018} is similar to \texttt{m00} in some respects, despite the fact that their accretion rate is lower by a factor of a few and their BH spin is much larger ($a_*=0.8$ vs $a_*=0$ in \texttt{m00}). They obtain a bolometric luminosity of $L \sim 2-3\,L_{\textrm{Edd}}$ when viewed nearly face-on and $L \sim L_{\textrm{Edd}}$ when viewed edge-on, just like \texttt{m00}. In addition, they too predict the presence of soft X-ray emission even for an edge-on observer. Note that they account for the absorption of soft X-rays by helium ions, which we do not model in this work (although the effect is roughly accounted for through our frequency-dependent opacity model).

One interesting point is that the BH spin of 0.8 in \citet{Dai2018} is much closer to what we used in model \texttt{m09} ($a_*=0.9$) than to model \texttt{m00} ($a_*=0$). Nevertheless, their model is more similar to \texttt{m00} in the sense that they do not find a relativistic jet just as in \texttt{m00}, whereas our model \texttt{m09} has a powerful and highly relativistic jet. One caution is that \citet{Dai2018} employed very low numerical resolution, viz., $128$ cells in $r$ covering the same range of radius as in our work, 64 cells in $\theta$ from $0$ to $\pi$, and 32 cells in $\phi$ for the full 0 to $2\pi$. Our models \texttt{m00} and \texttt{m09} have much larger resolution, $320$ cells in $r$, 192 cells in $\theta$ and 32 cells in $\phi$ for the range 0 to $\pi$. It is possible that their low resolution prevented the formation of a strong jet. It would be of interest to carry out high resolution simulations for other intermediate spin values.

The spectrum of the jetted TDE model \texttt{m09} is very different from the spectra of the other models, as is evident from Figure \ref{fig:fig12}. This model emits strongly in high energy bands (X-ray and $\gamma$-ray). The hard radiation is strongly beamed along the axis, and the apparent luminosity for on-axis observers is $\gtrsim 10^4 L_{\rm{Edd}}$. The luminosity falls off for larger viewing angles, but even at $\theta=90^\circ$, the luminosity is still $\gtrsim 30 L_{\rm{Edd}}$. On the other hand, the thermal emission from the torus in the NUV is largely independent of the viewing angle. It is sub-dominant in most direction, and becomes comparable to the hard emission only for edge-on observers. Note that there are no non-thermal electrons in the models studied in this paper. Even the hard radiation is produced by thermal gas. The large amount of X-ray and gamma-ray emission from \texttt{m09} is because (i) the gas in the jet is very hot, so the intrinsic emission is hard, and (ii) any soft radiation present tends to be Compton-upscattered by the same hot electrons. The soft radiation for Comptonization is provided from the funnel wall. Thermal synchrotron emission produces a far infrared peak, as seen in the bottom panel of Figure \ref{fig:fig13}.

\begin{figure*}
  \centering
  \includegraphics[width=0.31\textwidth]{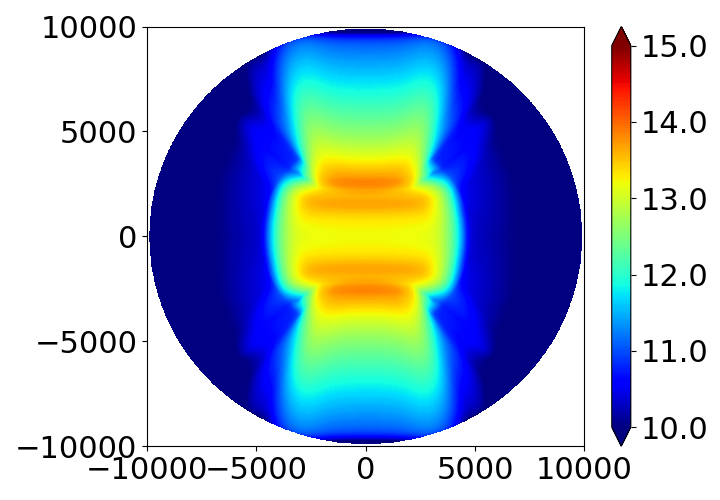}
  \includegraphics[width=0.31\textwidth]{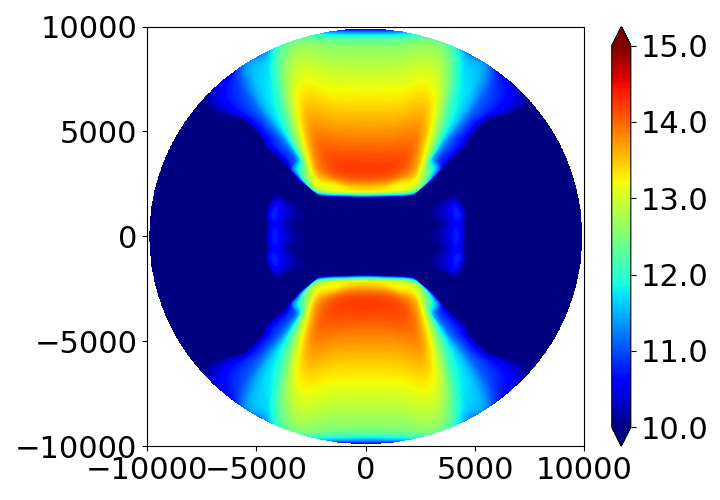}
  \includegraphics[width=0.31\textwidth]{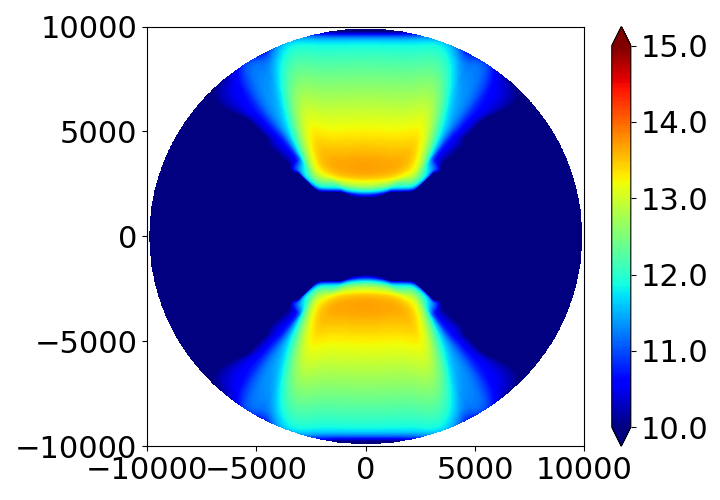}
  \caption{Images of model \texttt{m09} when viewed edge-on ($\theta=90^\circ$) in the optical/UV band (0.002-0.009 keV, left), X-ray band (0.2-10 keV, middle), and $\gamma$-ray band (15-150 keV, right).}
    \label{fig:fig14}
\end{figure*}

In Figure \ref{fig:fig14} we show images of model \texttt{m09} for an observer at $90^\circ$ viewing angle. The optical/UV emission (left panel) is partly from the torus and partly from the outer boundary of the jet, the two contributions being roughly equal. In the X-ray band (middle panel), the image is completely dominated by the jet. Notice the dramatic change in intensity and width of the X-ray emitting region when compared to the equivalent panel in Figure \ref{fig:fig13}. In the $\gamma$-ray band, most of the photons again appear to originate from the region of the jet that is outside the torus photosphere. Again, there is an enormous difference between the jetted \texttt{m09} model in Figure \ref{fig:fig14} and the non-jetted \texttt{s00} model in Figure \ref{fig:fig12}.

\subsection{Detection Limits}

\begin{figure}
   	\includegraphics[width=\columnwidth]{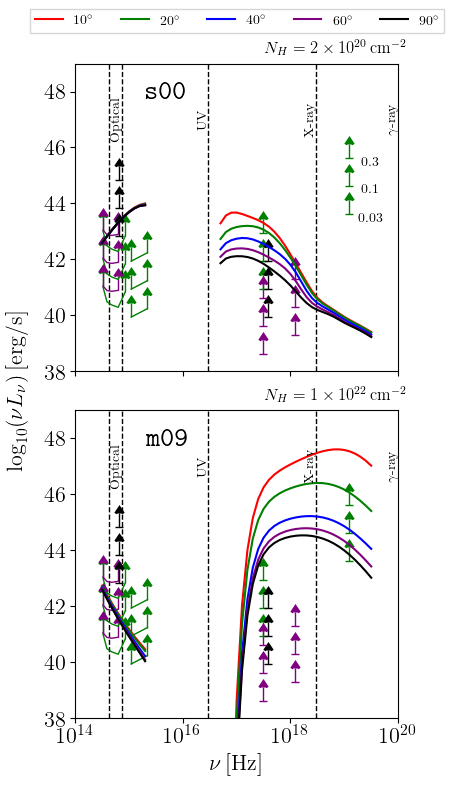}
    \caption{Minimum detectable luminosities of various telescopes (shown by arrows) compared with the predicted extincted luminosities for various observing angles (solid color lines, see legend at top) for models \texttt{s00} and \texttt{m09}. In the optical band, limits are shown for SDSS \textit{ugriz} (green arrows), ASAS-SN (black), Pan-STARRS1 (purple), in UV the limits correspond to GALEX DIS (green), in soft X-ray the limits are for \textit{Swift} XRT (green), \textit{Chandra} ACIS (black), \textit{XMM-Newton} (purple), and in hard X-ray/$\gamma$-ray the limits are for \textit{Swift} BAT (green). In each case, three arrows are shown, the lowest corresponding to a source located at $z=0.03$, the middle corresponding to $z=0.1$, and the uppermost to $z=0.3$. The ordering is indicated on the right hand side for the \textit{Swift} BAT band in the top panel. A hydrogen column density of $N_H = 2\times10^{20}~{\rm cm^{-2}}$ was used for the non-jetted model \texttt{s00} (upper panel) and $N_H = 1\times10^{22}~{\rm cm^{-2}}$ for the jetted model \texttt{m09} (lower panel). }
\label{fig:fig15}
\end{figure}

In this subsection we compare our models with detection limits for various state of the art telescopes. Namely, we include limits for the Sloan Digital Sky Survey (SDSS, \citealt{York2000}), Pan-STARRS1 \citep{Kaiser2002}, ASAS-SN \citep{Shappee2014}, GALEX Deep Imaging Survey (DIS, \citealt{Martin2005}), \textit{Swift} X-ray Telescope (XRT, \citealt{Burrows2005}), \textit{Chandra} Advanced CCD Imaging Spectrometer (ACIS, \citealt{Burke1997}), \textit{XMM-Newton} \citep{Jansen2001}, and \textit{Swift} Burst Alert Telescope (BAT, \citealt{Barthelmy2005}). Motivated by the distances of many of the previously discovered TDEs \citep{Komossa2015,Auchettl2017}, we consider sources in the redshift range $z=0.03-0.3$. We compute $4-5\sigma$ detection limits for each instrument in the relevant band using the typical exposure time. The assumed exposure times are 55 s (SDSS), 114-240 s (Pan-STARRS1), 30 ks (GALEX DIS), $10^4$ s (\textit{Swift} XRT, \textit{Chandra}, \textit{XMM-Newton}), and $10^6$ s (\textit{Swift} BAT). For ASAS-SN, we use a $V$ band limiting magnitude of $17$ mag.

In Figure \ref{fig:fig15}, we compare extincted band luminosities computed from the model spectra of \texttt{s00} and \texttt{m09} with detection limits for a point source. Since the hydrogen column density in jetted TDEs has been observed to be significantly enhanced in comparison to the average non-jetted TDE (e.g. see \citealt{Auchettl2017}), we consider detection limits using a different hydrogen column density ($N_H$) for the two classes. Based on the reported column densities, we assume $N_H = 2\times10^{20}$ ${\rm{cm^{-2}}}$ for the three non-jetted TDE models, \texttt{s00, s09, m00}, and $N_H = 1\times10^{22}$ ${\rm{cm^{-2}}}$ for the jetted TDE model \texttt{m09}. We describe the method we use to compute extinction in Appendix \ref{sec:appB}. We also examine detection limits for \texttt{m00} and \texttt{s09} in detail, as well as for all four models for the case where the column has an intermediate value, $N_H = 2\times10^{21}$ ${\rm{cm^{-2}}}$, and show the results in Figures \ref{fig:figC8}-\ref{fig:figC10}.

\subsubsection{Non-jetted TDEs}
For model \texttt{s00}, the optical/UV band luminosities are essentially the same for all viewing angles since the quasi-spherical torus is the source of this emission. As the upper panel in Figure \ref{fig:fig15} indicates, for both SDSS and Pan-STARRS1, the optical emission in \texttt{s00} is bright enough to be detectable out to $z \gtrsim 0.3$. In the case of ASAS-SN, only for $z\lesssim0.03$ will \texttt{s00} be detectable. In the UV we predict that, for the assumed low hydrogen column density of $N_H = 2\times10^{20}~{\rm cm^{-2}}$, the NUV emission is detectable by GALEX well beyond $z=0.3$. For \textit{Swift} XRT, the soft X-ray emission is detectable for all viewing angles at $z=0.1$, but only for more pole-on viewing angles at $z=0.3$. Even in the latter case, follow-up observations with more sensitive instruments (e.g. \textit{Chandra} or \textit{XMM-Newton}) should pick up the soft X-ray emission for any viewing angle based on our model spectra. The $\gamma$-ray emission in \texttt{s00} is extremely weak and would not be detected as a GRB by {\it Swift} BAT. The results are identical for model \texttt{s09} (see Figure \ref{fig:figC8}).

In the case of model \texttt{m00}, since the outflow is largely optically thick, the spectrum is nearly identical at all viewing angles (see Figure \ref{fig:figC8}). In fact, the spectrum resembles that of \texttt{s00} and \texttt{s09} at the equatorial plane. As such, \texttt{m00} meets the same detection limits as discussed above, except that the X-ray band luminosity is much lower for pole-on viewing angles.

For completeness, we also considered the case when the column density is an order of magnitude larger: $N_H = 2\times10^{21}$ ${\rm{cm^{-2}}}$); however, our conclusions regarding the detection limits do not change much, since the extinction is less than one magnitude in each band (see Figures \ref{fig:figC9} and \ref{fig:figC10}).

It is worth noting that the upper limit on the distance out to which the non-jetted models, \texttt{s00}, \texttt{s09}, and \texttt{m00}, can be detected in the $V$ band (with the assumed low column density) match well the redshifts of previous ASAS-SN detections of TDEs with a BH mass $\lesssim 10^7 \, M_\odot$. ASASSN-14ae \citep{Holoien2014}, ASASSN-14li \citep{Holoien2016a}, and ASASSN-15oi \citep{Holoien2016b} are all quite nearby TDEs with redshifts between $z=0.0206-0.0484$. Indeed, the analysis of each of these objects suggests the optical/UV emission originates from a thermal source of $T\sim 10^4$\,K, much like in our models. We do not consider ASASSN-15lh \citep{Dong2016,Leloudas2016,Godoy-Rivera2017} in this comparison as the BH mass is too large for the peak accretion rate to be super-Eddington.

The detection limits described above are considered only for the emission from the accretion flow. 
Based on this analysis, we do not predict that viewing angle dependence alone can lead to a non-detection of either optical/UV or X-ray emission \textit{at peak} accretion for a $10^6\, M_\odot$ BH. 
The previously discovered non-jetted TDEs have all had relatively small column densities \citep{Auchettl2017}, so it is unlikely that extinction is enough to explain the non-detection of optical/UV TDEs in the X-ray, though it is possible that reprocessing of emission by debris at larger radii could lead to significant absorption in the X-ray \citep{Guillochon2014}.

\citet{Roth2016} find that TDE emission in the optical through X-ray is sensitive to the mass of the envelope of absorbing material (e.g. the large scale torus in our model) and the luminosity of the source. If the bolometric luminosity becomes low enough, they predict the formation of a helium recombination front which can completely absorb the X-ray photons. While this model provides some intuition for why some TDEs show no X-rays, the geometry of our models are quite different. They consider a spherical symmetry whereas our models (save \texttt{m00}) have an optically thin funnel of low density gas that emits soft X-rays that can be detected even by an edge-on observer. In addition, the model presented by \citet{Dai2018} shows that, even with ions included in the radiative transfer, the hot wind and outflow will produce luminous, super-Eddington X-ray emission that will be detected by an edge-on observer. It is important to note that \citet{Dai2018} and our work only explore the case of a near solar mass star being disrupted. For higher mass stars (and thus more massive envelopes), the additional absorbers in the torus could potentially absorb out the soft X-rays even at small angles (i.e. $\theta < 30^\circ$) where much of the X-ray emission is escaping in model \texttt{s00}. This could explain the observation of veiled TDEs in the context of a super-Eddington accretion disk.

\subsubsection{Jetted TDEs}

We discuss the spectrum of our jetted TDE model \texttt{m09} assuming a large column density, $N_H = 1\times10^{22}$ ${\rm{cm^{-2}}}$, as has been seen in the jetted TDEs discovered so far. The emission in the optical/UV in this model (lower panel in Figure~\ref{fig:fig15}) is somewhat enhanced relative to the non-jetted TDE models because, apart from radiation from the torus, there is also some optical/UV radiation from the outflow. As a result, despite the larger column, this model is detectable in the optical and NUV up to $z \sim 0.1$. The FUV is, however, undetectable. The jet emission is very luminous, so the X-ray emission is detectable by all instruments at all viewing angles, even for $z>0.3$. 

We also examine the case of a lower column density, as in J2058 and J1112 (see Figure \ref{fig:figC9}). Extinction of high energy photons is negligible, so we find the same detection limits in the X-ray and $\gamma$-ray bands. Predictably, the extinction of optical and UV photons is much less and our models indicate that an object like J2058 (or J1112) should appear in the optical and NUV for the instruments considered.

For both cases, the $\gamma$-ray emission from model \texttt{m09} is detectable by \textit{Swift} BAT up to $z\gtrsim 0.03-0.3$ depending on the orientation of the jet. Interestingly, our analysis suggests that jetted TDEs at redshifts larger than $z=0.3$ will only be detected for nearly face-on observers ($\theta \lesssim 20^\circ$). All three of the proposed jetted TDEs have been discovered at quite large distances, $0.358\leq z \leq 1.1853$ \citep{Bloom2011,Burrows2011,Zauderer2011,Cenko2012,Brown2015}. J1644 is the closest jetted TDE at $z=0.358$ and was likely observed face-on given the strongly beamed emission. Given the rather large distance of J1112 ($z=0.89$) and J2058 ($z=1.1853$) our models suggest near face-on observations of these events as well.

\subsection{Comparison with observational properties of TDEs}

\begin{figure}
	\includegraphics[width = \columnwidth]{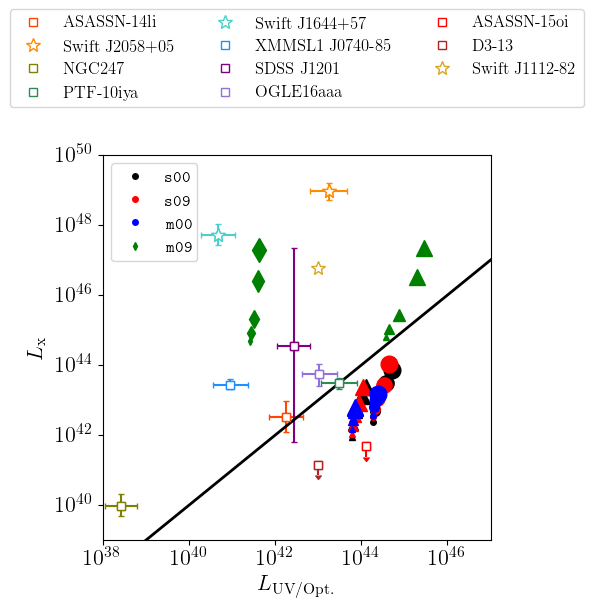}
    \caption{Plot of the luminosity in the x-ray band (0.3-10 keV) versus the luminosity in the UV/optical band (0.002-0.1 keV). Results from the four simulated models (coded by color) are shown for $N_H = 2\times10^{20}$ ${\rm{cm^{-2}}}$ (filled circles), $N_H = 2\times10^{21}$ ${\rm{cm^{-2}}}$ (filled triangles), and $N_H = 1\times10^{22}$ ${\rm{cm^{-2}}}$ (filled diamonds). In each set, the largest point corresponds to an observer aligned with the jet axis while the smallest corresponds to an observer viewing the system edge-on. Observational data \citep{Gezari2009,Cenko2012,Holoien2016b,Auchettl2017} are plotted as open squares (non-jetted TDEs) and open stars (jetted TDEs).}
    \label{fig:fig16}
\end{figure}

\begin{figure}
	\includegraphics[width = \columnwidth]{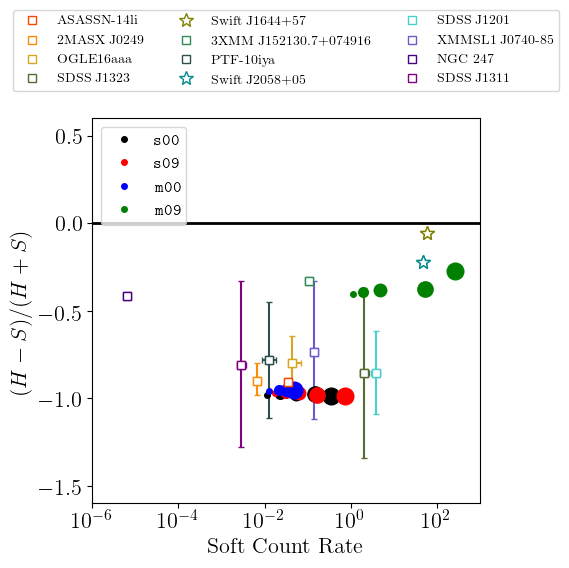}
    \caption{Plot of the X-ray hardness ratio versus the count rate in the soft X-ray band for a source at redshift 0.1. Results for the four simulated models (coded by color) are shown as filled circles. In each set, the largest point corresponds to an observer aligned with the jet axis while the smallest corresponds to an observer viewing the system edge-on. Observational data from \citet{Auchettl2017}, rescaled to $z=0.1$, are plotted as open squares (non-jetted TDEs) and open stars (jetted TDEs).}
    \label{fig:fig17}
\end{figure}


In the thorough analysis of the catalog of TDEs carried out by \citet{Auchettl2017}, the observational characteristics of the different classes of TDEs are examined in great detail. Of particular interest to us is the separation of TDE emission properties at peak. They find that at peak (a) jetted TDEs have a relatively hard X-ray spectrum, producing almost equal counts in the soft and hard X-ray bands, while non-jetted TDEs tend to be softer, (b) non-jetted TDEs radiate nearly equal amounts of energy in the X-ray band as they do in the UV/Optical bands, while the jetted TDEs emit much more energy in X-rays, with an X-ray band luminosity up to nearly 5-6 orders of magnitude higher than the UV/Optical, and (c) jetted and non-jetted TDEs self-separate when comparing the hard (2-10 keV) and soft (0.3-2 keV) count rates. Here we use the same bands and definitions as in \citet{Auchettl2017} to compare the spectra of our simulation-based jetted and non-jetted models with observations.

In the computations, we assume that the source is located at redshift $z=0.1$. In order to compute count rates, we assume 100\% of the photons are detected and use the effective area of the \textit{Swift} XRT.

\subsubsection{Optical/UV and X-ray Emission}

In Figure \ref{fig:fig16}, we compare the X-ray luminosity (0.3-10 keV emission) and the UV/Optical luminosity (0.002-0.1 keV emission) for each model. We show observational data \citep{Gezari2009,Cenko2012,Holoien2016b,Auchettl2017} for several TDEs at peak as open squares (non-jetted TDEs) and open stars (jetted TDEs). Note that for D3-13 and ASAS-SN 15oi upper limits of the peak X-ray luminosity are provided as the peak X-ray emission in both events is comparable to the host galaxy's emission in a quiescent state. For comparison, we show our model results for non-jetted TDEs for $N_H = 2\times10^{20}$ ${\rm{cm^{-2}}}$ (filled circles) and $N_H = 2\times10^{21}$ ${\rm{cm^{-2}}}$ (filled triangles), and for jetted TDEs for $N_H = 2\times10^{21}$ ${\rm{cm^{-2}}}$ (triangles) and $N_H = 1\times10^{22}$ ${\rm{cm^{-2}}}$ (diamonds). The sizes of the points decrease as the viewing angle increases, i.e., the largest symbols correspond to viewing down the jet axis and the smallest symbols to viewing edge-on. 

For $N_H = 2\times10^{20}\,{\rm{cm^{-2}}}$, we find that the non-jetted TDE models, \texttt{s00, s09, m00}, have X-ray luminosities roughly consistent with observations (compare open squares  and filled circles in Figure \ref{fig:fig16}). However, the models predict more optical/UV luminosity than observed in TDEs that have well constrained X-ray emission. The discrepancy is more than an order of magnitude, which is difficult to understand. A characteristic (and unavoidable) feature of our super-Eddington models is that they will emit thermal Optical/UV radiation with a luminosity of around Eddington, i.e., $10^{44}~{\rm erg\,s^{-1}}$. This statement should be true for any viewing angle. Assuming the observed systems are super-Eddington accretors, two possible explanations are (i) the BH masses are much smaller than the $10^6M_\odot$ mass used in our simulations, and (ii) the extinction columns are much larger than $N_H = 2\times10^{20}\,{\rm{cm^{-2}}}$. Neither option is very likely. For instance, the estimated BH masses are, if at all, larger than $10^6M_\odot$ for many systems. Also, the hydrogen column densities of non-jetted TDEs have thus far not been very large. \citet{Auchettl2017} report values generally around $N_H = 10^{20-21}\,{\rm{cm^{-2}}}$ with most of the well constrained TDEs having $N_H\sim 10^{20}\,{\rm{cm^{-2}}}$. Given \citet{Auchettl2017} obtain the NUV/FUV component of the spectrum by fitting a decaying power law to the data, another possibility is that the reported luminosities underestimate the NUV/FUV emission. For instance, note that our model non-jetted TDE spectra (Figure \ref{fig:fig12}) have a second peak in the FUV which has a significant luminosity at near face-on observing angles. It is important to note that FUV emission from TDEs has yet to be observed, and our models produce significant FUV emission from the outflow. It is also possible that the peak emission of certain TDEs was not picked up in transient surveys due to insufficient cadence.

In the case of the jetted TDE model \texttt{m09}, the model predictions shown in Figure \ref{fig:fig16} agree fairly well with the observations when we use a large column of $N_H = 1\times10^{22}$ ${\rm{cm^{-2}}}$. The agreement with J1644 is particularly good, provided we accept the general assumption that the source was observed at a small inclination angle. In the case of J1112 and J2058, we would obtain reasonable agreement with a slightly smaller $N_H$, but the estimated $N_H$ is substantially smaller ($2\times 10^{21}\,{\rm{cm^{-2}}}$). Another issue is that J2058 had an X-ray luminosity of $10^{49}~{\rm erg\,s^{-1}}$, whereas model \texttt{m09} barely reaches $10^{48}~{\rm erg\,s^{-1}}$ for a face-on observer. This is not a serious discrepancy because we have considered only a single fiducial model here. By changing the BH mass, BH spin or mass accretion rate, it ought to be possible to obtain the required luminosity.

\subsubsection{X-ray Spectral Hardness}

The hardness ratio HR is defined as $(H-S)/(H+S)$, where $H$ is the count rate in the hard (2-10 keV) band and $S$ is the count rate in the soft (0.3-2 keV) band.  In Figure \ref{fig:fig17}, we plot HR as a function of the count rate in the soft band. We again include observational data for several confirmed TDEs at peak from \citep{Auchettl2017} and compare them against predictions of our models. Note that we do not include extinction when computing the hardness ratio for our models, as the spectral hardness is likely uncorrelated with the column density \citep{Auchettl2017}. Both data and models are scaled for source redshift of 0.1. For the observational data, we only include those events that are classified as the disruption of a star by a SMBH.

The non-jetted models become harder as the viewing angle increases towards the equatorial plane and they populate HR values between -1 and -0.9. The non-jetted TDE data reasonably match the models with the exception of 3XMM J152130.7+074916 (3XMM hereafter) and NGC 247. Interestingly, model \texttt{m09} has a slightly softer spectrum than the estimates from \citet{Auchettl2017} for the jetted TDEs J1644 and J2058. This is true even for a nearly face-on observer. Despite these minor caveats, the qualitative agreement is quite good in that there is a clear separation in spectral hardness between the jetted and non-jetted TDE model spectra.

3XMM and NGC 247 appear to be quite hard compared to the rest of the non-jetted TDEs, which is puzzling. Both 3XMM and NGC 247 are classified as \textit{likely TDEs} in \citet{Auchettl2017}, but they clearly stand out from most other TDEs in the literature. 3XMM is classified as a TDE partly because of its transient nature, but the BH mass inferred from the X-ray data is small compared to other TDEs ($M_{\rm{BH}} \sim 10^5 - 10^6\,M_\odot$) and it has not been well constrained to the center of the host galaxy \citep{Lin2015}. NGC 247 is both less luminous than the typical TDE and is significantly harder. \citet{Feng2015} note that the lack of data on NGC 247 for the 3 years prior to the transient event may imply that the transient actually appeared 1-3 years before being detected. This might imply that the accretion rate is not near the peak, and thus the system is possibly not comparable to our models. 

\subsection{Comparison with \textit{Swift} J1644+57}

\subsubsection{Emission Properties}

The observational characteristics of the jetted TDE J1644 are qualitatively similar to the spectra we compute for model \texttt{m09}. In particular, the model spectrum extends to the $\gamma$-ray band, consistent with the \textit{Swift}-BAT detection. The emission in the GeV-TeV band is weak or absent, consistent with \textit{Fermi} LAT and Veritas upper limits \citep{Burrows2011}. \citet{Aliu2011} and \citet{Aleksic2013} find that J1644 does not show $\gamma$-ray emission at frequencies greater than $10^{25}$ Hz, similar to our model. In fact, the spectrum of \texttt{m09} cuts off quite abruptly at $10^{23}$\,Hz. The peak luminosity observed by an on-axis observer is roughly $10^{48}$ ${\rm erg\,s^{-1}}$ for the model, within a factor of a few of the observed luminosity.

The evolution of J1644 has been followed in detail from radio to hard X-ray bands. The observations suggest that the jet is responsible for the X-ray emission while the interaction between the jet and surrounding medium leads to a shock which produces non-thermal synchrotron emission in the radio bands \citep{Bloom2011,Metzger2012,Berger2012,Zauderer2013}. \citet{Crumley2016} considered several radiation mechanisms and find that external IC (within the jet, as in \texttt{m09}) or magnetic reconnection in a Poynting dominated jet are favorable mechanisms. In addition, \citet{Auchettl2017} claim that the IR emission is well described by optically thin thermal synchrotron emission while emission in higher energy bands (up to the UV) can be described by thermal blackbody emission. Our model (\texttt{m09}) generally matches predictions made in the literature (see Figure \ref{fig:fig12} and discussion in Section 5.2) with the exception of the radio component which is usually interpreted in terms of a shock as the jet interacts with the external medium. Note that the earliest radio data come from $\delta t \approx 4$ days after the BAT detection while our simulations were run for a total time of only $\sim 1$ day. Also, our simulations do not extend to a large enough radius, nor do they attempt to model a realistic external medium, so the lack of emission from an external shock in the models is to be expected.

\citet{Tchekhovskoy2014} proposed that the large X-ray luminosity and highly variable light curve of J1644 could be explained simply by assuming that a strong magnetic flux was present in this system and powered the jet. They suggested that the BH and disk spin axes were likely misaligned when the accretion disk first formed, which would lead to an initially precessing jet that later becomes aligned with the BH spin. This would produce the observed variability and explain the late-time radio emission. We agree with \citet{Tchekhovskoy2014} that the J1644 transient was likely powered by a MAD TDE accretion disk around a rapidly spinning BH; however, we leave considerations of quasi-periodic oscillations in the X-ray emission to a future analysis.

\subsubsection{Jet Structure}
With the abundance of observational data on J1644, the jet structure has been examined in several studies.  The transient exhibited strongly beamed radiation from an ultra-relativistic jet with $\theta_j = 1/\Gamma_j \sim 0.1 \approx 6^\circ$ \citep{Metzger2012}. Follow-up radio observations presented by \citet{Berger2012} showed that the radio emission re-brightened well after the initial burst of emission. This suggested that the model used by \citet{Metzger2012} of a single $\Gamma$ component blast-wave was insufficient and that there might be a slower moving component that shocks later than the faster moving jet core.

A two component model has been employed in several studies to model the jet structure and emission \citep{Wang2014,Mimica2015,Liu2015}. Each of these works suggests that the emission is best explained by a two component jet that is separated in velocity space.

The jet structure produced by model \texttt{m09} is similar to the best fit models of \citet{Mimica2015}. The central jet in our simulated model has a relatively large Lorentz factor ($\Gamma \sim 2-6$, compared with $\Gamma \sim 10$ in \citealt{Mimica2015}), and it is surrounded by a slower moving, mildly relativistic outflow in a sheath ($\Gamma \lesssim 2$ in our model and in \citealt{Mimica2015}). We also find a similar opening angle for the sheath as the value they report: $\theta_{j,s}\sim 29^\circ$.

We compute the total kinetic energy contained in the core and sheath directly from the \textsc{KORAL} data on model \texttt{m09}, using cells over the radius range $1000 r_g < r < 5000 r_g$. This range of radii is well inside the jet head and spans the region where the jet is roughly conical. We integrate the kinetic energy over the core and sheath regions to obtain the net kinetic energy in the two regions, as described in equation (\ref{eq:Ek}) using the angular extent of each region described in Section \ref{sec:sec442}. We compute the ratio of kinetic energy in the sheath versus the core as $R_{\rm{k,jet}} = E_{\rm{k,sheath}}/E_{\rm{k,core}}$. Using our previously estimated core opening angle of $\theta_{j,c}\sim 15^\circ$, we find that the above ratio is only a bit greater than unity: $R_{\rm{k,jet}} \sim 2$. If we instead use the jet opening angle reported by \citet{Metzger2012}, we find $R_{\rm{k,jet}} \sim 13$.

An explicit assumption made by \citet{Mimica2015} is that the core and sheath of the jet contribute significantly to the early X-ray emission. The X-ray emission from the jet in \texttt{m09} clearly has significant contributions from regions extending all the way to the edge of the jet (see Figure \ref{fig:fig14}). If jetted TDEs are indeed MAD and super-Eddington, the early X-ray emission would be expected to originate from both the relativistic and mildly relativistic outflow.

\section{Discussion and Conclusions}
We used the general relativistic radiation MHD (GRRMHD) code \textsc{KORAL} to carry out numerical simulations of a super-Eddington accretion disk that forms after the disruption of a $1M_\odot$ star by a $10^6 \, M_\odot$ SMBH. We ran four simulations with parameters designed to explore how the dynamics and radiative properties of the accretion flow depend on the BH spin $a_*$ and magnetic field strength (SANE vs MAD). We initialized the models with a weakly bound, constant angular momentum torus of mass $0.17\,M_\odot$, which is in the mass range expected for typical TDEs. The resulting mass accretion rate is around 100 times Eddington, as appropriate for the peak of a TDE transient. One of our models, \texttt{m09}, is to our knowledge the first GRRMHD simulation of a jetted TDE. 

We post-processed the output from the four simulations using the radiative transfer code \textsc{HEROIC}, and computed spectra and images as a function of the viewing angle of a distant observer. We then carried out a comprehensive comparison of the model spectra with TDE observations.

Three of our models, \texttt{s00, s09, m00}, agree well with observations of non-jetted TDEs, while the fourth model, \texttt{m09}, closely resembles jetted TDEs. The latter model has a rapidly spinning BH ($a_*=0.9$) and develops a strong dipolar magnetic field at the BH horizon, i.e., it is in the MAD state. Evidently, both rapid BH spin and MAD accretion are necessary to produce a jetted TDE.

The three non-jetted TDE models, \texttt{s00}, \texttt{s09}, \texttt{m00}, are highly inefficient, with a radiative efficiency $\eta_r \lesssim 1$\%. Including all forms of energy (radiation, kinetic, magnetic), however, these systems are somewhat more efficient, with $\eta_t\sim 2-7$\%. The models \texttt{s00} and \texttt{s09} both launch a wind and `jet', though the latter is not a true relativistic jet but is more accurately described as a radiation driven outflow. The energy carried by the jet is not very large, $\eta_{t,\rm{jet}}\sim 0.5-1.7$\%, and there is roughly an equal amount in the wind. The MAD, non-jetted model \texttt{m00} on the other hand may be thought of as a wind dominated accretion disk since it has $\eta_{t,\rm{jet}} \ll \eta_{t,\rm{wind}}$.

The jetted model \texttt{m09} is very different from the other three models. It is characterized by rather high efficiencies, with $\eta_r \sim 64$\%, and $\eta_t \approx 100$\%. The very large efficiency of this model is because the accretion flow extracts a large amount of spin energy from the BH. Almost all of this energy goes into the jet. Model \texttt{m09} also launches a wind, but the wind efficiency is much smaller, $\eta_{t,\rm{jet}} \gg \eta_{t,\rm{wind}}$. For comparison, \citet{Dai2018} presented a MAD, $a_*=0.8$, $\dot{M} = 15\dot{M}_{\rm{Edd}}$ TDE accretion disk with $\eta_r \sim 2.7$\%, $\eta_{t,\rm{jet}} \sim 20$\%, and $\eta_t \sim 43$\% (their accretion rate was  several times smaller than in our models). While their model did not produce a relativistic jet, as did \texttt{m09}, it may be taken as an intermediate case between our models \texttt{m00} (a low spin, MAD model) and \texttt{m09} (a near extremal spin, MAD model). The effect of increasing the spin of the BH clearly leads to the injection of more energy into the jet and a higher radiative efficiency; however, the lack of a jet in their model suggests $a_* > 0.8$ is needed to produce a truly relativistic jet. We should note that it is possible that the lack of an ultra relativistic jet in \citet{Dai2018} is due to inadequate numerical resolution.

We computed model spectra from our simulations and compared them with observational data on jetted and non-jetted TDEs. We found surprisingly good qualitative agreement. We did note some quantitative discrepancies, which call for more detailed study, but the overall conclusion is that the super-Eddington models described here provide a promising explanation of TDE phenomenology, at least near the peak of these outbursts.

The spectra of the three non-jetted models have many similarities with observations of non-jetted TDEs. The model spectra are double peaked, with emission from the torus producing optical/UV emission at $T_r\sim 10^{4.4}$ K, while the funnel walls and heated wind produce a component peaking in the UV/soft X-ray at $T_r\sim 10^{5-6}$ K. The radiation driven outflow is accelerated to high velocities with $v/c \sim 0.3-0.6$ ($\Gamma\sim 1.1-1.2$) in the jet, depending on the spin of the BH. The presence of an ultrafast outflow has been inferred in the non-jetted TDE ASASN-14li in both X-rays and radio \citep{Alexander2016,Kara2018}. 

The optical/UV luminosity of the non-jetted models is around Eddington, which for our $10^6M_\odot$ BH corresponds to $\sim10^{44} {\rm erg\,s^{-1}}$. The observed luminosity is nearly independent of the viewing angle. The X-ray luminosity is of the order of $10^{42-44}$ ${\rm erg\,s^{-1}}$, varying from the upper end for face-on observers to the lower end for edge-on observers. The X-ray luminosity of model \texttt{m00} is nearly independent of viewing angle and is near the lower end of the range. The X-ray spectral shapes (hardness ratios) are broadly consistent with observations. 

An interesting point is that the observed X-rays in the non-jetted TDE models do not come from the `base of the jet' or a `corona above the accretion disk' near the BH, as usually assumed. Rather, they come from the jet and wind at radii of several thousand $r_g$, where the outflowing material emerges outside the optically thick torus. This radiation is visible for all viewing angles. \citet{Dai2018} reached a similar conclusion, viz., that X-ray photons from polar regions can reach even edge-on observers.

Comparing our model spectra with the detection limits of various telescopes and surveys we find that, in the absence of a significant hydrogen column, our non-jetted models should be detectable in optical, UV, and soft X-rays, regardless of viewing angle. The models thus do not explain why some optical/UV non-jetted TDEs are not detected in X-rays. This is an area of discrepancy between the models and observations. Furthermore, this result disagrees with the idea, proposed by \citet{Dai2018}, that the viewing angle can explain the different TDE classes if TDE accretion disks are geometrically thick. While the ratio between X-ray and optical/UV luminosity does decrease with increasing viewing angle, there is still a detectable X-ray flux even for an edge-on observer. This statement is true even for the spectra presented by \citet{Dai2018}, so in both models the X-ray emission ought to be detectable regardless of viewing angle.

Turning to model \texttt{m09}, we find that a MAD accretion disk around a rapidly spinning BH produces a powerful jet that can reproduce several features observed in the jetted TDE J1644, in agreement with the proposal of \citet{Tchekhovskoy2014}. The model has a highly relativistic jet with $\Gamma \sim 6$, which is powered by spin energy from the BH, presumably via the \citet{Blandford1977} mechanism. The core of the relativistic jet has an opening angle of $\theta \sim 15^\circ$ and there is a mildly relativistic sheath at angles $\theta \sim 15-30^\circ$. The core and sheath carry nearly equal amounts of energy. 

The core-sheath structure of the jet is in agreement with models of the radio emission from J1644, which indicate that a single component cannot explain the late-time radio emission. However, models in the literature differ widely in the properties of the two regions. \citet{Wang2014} predicted that the sheath carries nearly twice as much energy as the core. On the other hand, the sheath could carry more than 25 times less kinetic energy than the core \citep{Mimica2015,Liu2015}. Our model \texttt{m09} has roughly equal amounts of energy in the core and the sheath.

Model \texttt{m09} also reproduces several features in the observed spectrum of J1644. The hard X-ray luminosity for a face-on observer is nearly $10^{48}$ ${\rm erg\,s^{-1}} \approx 10^4L_{\rm Edd}$, within a factor of a few of the luminosity observed at early times in J1644. The X-ray spectrum of model \texttt{m09}  is much harder than that of non-jetted models, in agreement with observations which show that jetted TDEs generally have harder spectra. The high velocity outflow launched by \texttt{m09} is consistent with velocities inferred from X-ray reverberation mapping of J1644 \citep{Kara2016}. Assuming an enhanced hydrogen column, as observed in J1644, the model is able to explain the unusually weak optical emission in this system. Finally, the model suggests that distant jetted TDEs ($z\gtrsim 0.3$) will be detected as GRBs only when viewed close to the jet axis, where relativistic beaming strongly enhances the observed luminosity. 

Between our two MAD models, \texttt{m00} and \texttt{m09}, and the MAD simulation presented in \citet{Dai2018}, a significant range of the spin parameter $a_*$ has been probed using GRRMHD simulations of MAD systems. These three simulations suggest that (i) the extraction of spin energy from the BH results in higher total efficiencies and injects a significant portion of this energy into the jet or wind, (ii)  the emission properties of both non-jetted and jetted TDEs can be reasonably described using a MAD accretion disk model, and (iii) a MAD disk with $a_* > 0.8$ is necessary to produce a jet. The third point follows from the fact that \citet{Dai2018} did not find a relativistic jet. It is a surprising result since the physics of jet acceleration as presently understood does not demand such large spins. We should caution that the lack of a relativistic jet in the \citet{Dai2018} model may be merely because of inadequate numerical resolution (their resolution was a factor of 2 to 3 lower along each  spatial dimension).

In summary, in this paper we have presented simulations of the accretion disk that forms from rapid circularization of TDE fall back material. We find good agreement between the dynamics and spectra of our simulated models and observational properties of both jetted and non-jetted TDEs. Our results confirm that a rapidly spinning BH with a MAD accretion disk is a likely explanation of jetted TDEs.

Before concluding, we discuss some caveats. The GRRMHD simulations with \textsc{KORAL} described here use the so-called M1 closure scheme \citep{Levermore1984} to model the radiation stress-energy tensor. The moment-based M1 method is more accurate than Eddington closure or simple diffusion (or even flux-limited diffusion), but it is nevertheless an approximation. Previous studies have shown that M1 closure, while perfectly adequate in most regions of the accretion disk, has difficulties in polar regions, where beams from the disk or funnel converge toward the axis. An artificial radiative viscosity mitigates the problem considerably \citep{Sadowski2015}, but it is likely that the treatment of radiation in the jet is still far from perfect. This is a potential issue for some of the quantitative results we report here regarding the speed and luminosity of the jet, since a part of the jet acceleration is from radiative driving.

The radiation post-processing code \textsc{HEROIC} does not suffer from this problem since it uses a large number of rays (162 rays in this study) and thus has ample angular information to handle physics near the axis. However, \textsc{HEROIC} does not solve for the gas dynamics -- it merely carries out a more accurate calculation of the gas thermodynamics and radiative transfer. Thus, at least in the matter of jet dynamics, the caveat of the previous paragraph remains.

One other minor point is that \textsc{HEROIC} uses a simple model for the frequency dependent opacity of atomic processes. In particular, it replaces atomic edges from bound-free transitions with a smoothed out opacity profile, and it does not include any line opacity. We do not think this is a serious issue, given the data presently available on TDEs, but we note that \citet{Dai2018} did include the opacity edge due to ionized helium in their calculations.

\section*{Acknowledgements}
We are grateful to James Guillochon for guidance throughout this work. We thank Aleksander S{\c a}dowski for assistance with the \textsc{KORAL} code, and Andrew Chael, Pawan Kumar, and especially Wenbin Lu and the referee for useful comments and suggestions. This work was supported in part by NSF grant AST-1816420, and made use of computational support from NSF via XSEDE resources (grant TG-AST080026N) and NASA via High-End Computing (HEC) resources. This work was carried out at the Black Hole Initiative at Harvard University, which is supported by a grant from the John Templeton Foundation.





\appendix

\section{Disk Initialization - Hydrostatic Rotating Disk with Power Law Angular Momentum} \label{sec:appA}
In this work, we use the power law angular momentum disk in hydrostatic equilibrium that was presented in \citet{Kato2004}. Here we briefly describe the model and how we initialize the torus in the \textsc{KORAL} code. 

For the model presented in \citet{Kato2004}, they use the pseudo-Newtonian potential described in \citet{Paczynsky1980}:
\begin{equation} \label{eq:eqA1}
  \phi = -\dfrac{GM}{(R-R_S)},
\end{equation}
where $R$ is the radius in polar coordinates, and $R_S$ is the Schwarzschild radius. A polytropic equation of state is assumed such that $p=K\rho^{1+1/n}$ and the angular momentum distribution of the disk is assumed to be a power law given by:
\begin{equation} \label{eq:eqA2}
  l(r,z) = l_0 \left(\dfrac{r}{r_0} \right)^a,
\end{equation}
where $r$ and $z$ are the cylindrical radius and height, and $l_0 = (GMr_0^3)^{1/2}/(r_0 - R_S)$. Here $r_0$ is simply a scale radius that sets the pressure and density maximum and $a$ is a constant. Under these assumptions, the condition for hydrostatic equilibrium combined with the polytropic equation of state yields a complete solution for the entire torus given the pressure ($p_0$) and density ($\rho_0$) at the characteristic radius $r_0$:
\begin{align}
  \rho &= \rho_0\left[ 1 - \dfrac{\gamma}{v_{s,0}^2}\dfrac{(\psi - \psi_0)}{n+1}\right]^{n}, \label{eq:eqA3} \\
  p &= \rho_0\dfrac{v_{s,0}^2}{\gamma}\left(\dfrac{\rho}{\rho_0}\right)^{1+1/n}, \label{eq:eqA4} 
\end{align}
where $\gamma$ is the adiabatic index (which we set to $4/3$ since the torus is radiation dominated which implies $n=3$), $v_s = \sqrt{\gamma p/\rho}$ is the sound speed of the gas, $\psi = \phi + \xi = -GM/(R-R_S) -l^2/2r^2(1-a)$ is the effective potential. Here $\xi$ is the centrifugal potential.

The Bernoulli parameter for the gas is given by the sum of the specific kinetic, potential, and internal energy. In the context of the power law angular momentum model employed here the gas is initially on a Keplerian orbit, so it may be expressed as:
\begin{equation} \label{eq:eqA5}
  \rm{Be} = (1-a)\xi + \phi + \psi_{\rm{int}}, 
\end{equation}
where $\psi_{\rm{int}} = \gamma p /(\gamma - 1)\rho$ is the internal potential. The condition of hydrostatic equilibrium satisfies the equation $\nabla(\xi + \phi + \psi_{\rm{int}}) = 0$, which implies:
\begin{equation} \label{eq:eqA6}
\xi + \phi + \psi_{\rm{int}} = {\rm{constant}}.
\end{equation}
Theoretical studies of TDE disks find that the gas comes in with roughly equal angular momentum. As such, we use a constant angular momentum model in this work. This implies that we should choose $a=0$. Under this condition, the Bernoulli parameter of the disk is also constant given equations (\ref{eq:eqA5}) and (\ref{eq:eqA6}).

To initialize the disk within the \textsc{KORAL} code, we specify the characteristic radius ($r_0$), maximum density ($\rho_0$), and initial gas temperature at the density maximum ($T_0$). We set the characteristic radius to be the circularization radius given by equation (\ref{eq:Rc}). The initial gas density effectively sets the accretion rate once the disk reaches a quasi-steady state, and the gas temperature is chosen such that the initial Bernoulli parameter of the torus matches the binding energy specified in equation (\ref{eq:depsilon}).

\begin{figure}
	\includegraphics[width=\columnwidth]{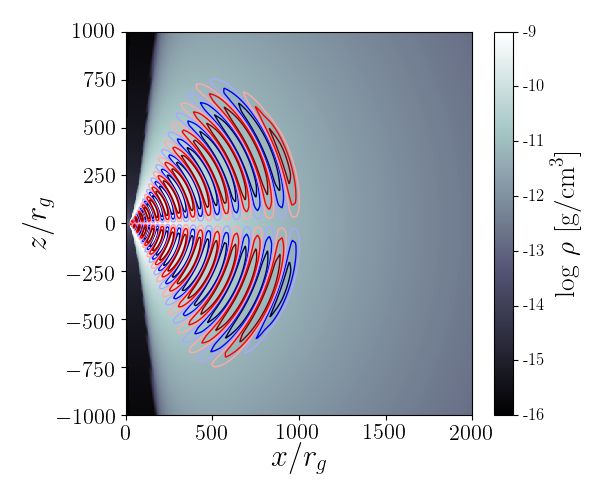}
   	\includegraphics[width=\columnwidth]{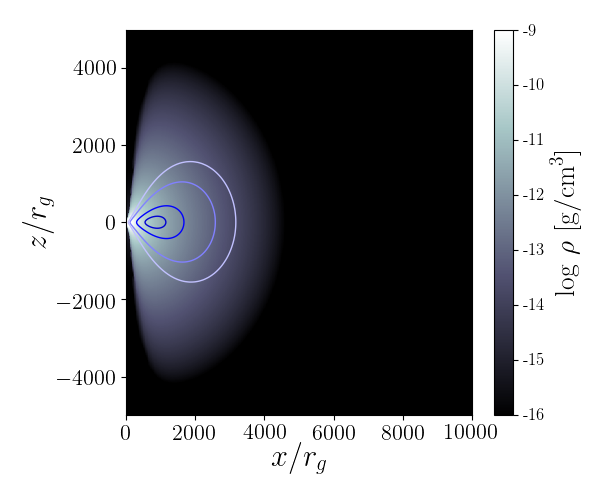}
    \caption{Initial magnetic field for the SANE (top) and MAD (bottom) models. For the SANE model the color of the contour indicates the sign of the magnetic field. We have zoomed in to better show the field structure. Note that the sign of the loop changes across the equatorial plane. For the MAD model we initialize the torus with a single poloidal loop of one sign. In this case we show the entire torus which extends to nearly $5000\,r_g$.}
    \label{fig:figA1}
\end{figure}

To achieve a MAD accretion disk, we initialize the magnetic field as a large dipolar field. This leads to the accumulation of magnetic field of only one polarization and the BH builds up a large magnetic flux quite rapidly. For the SANE models, we initialize the disk with multiple loops of alternating polarity. This prevents the build up of magnetic flux since the field cancels out. We show the gas density and field lines of the initial state of the MAD and SANE models in Figure \ref{fig:figA1}.

\section{Effect of Extinction} \label{sec:appB}
For the purposes of comparing the model spectra with observational results, we include the effects of extinction in the optical, UV, and X-ray (0.1-10 keV) bands. For the X-ray bands, we use the interstellar medium (ISM) particle cross section as a function of energy as calculated in \citet{Wilms2000}. The extinction is higher in the soft X-ray than in the hard X-ray and is given by:
\begin{equation} \label{eq:eqB1}
  I_{\rm{obs}}(E) = I_0(E) \exp[-\sigma_{\rm{ISM}}(E)N_H]
\end{equation}
where $E$ is the photon energy, $I_0$ is the initial intensity, $I_{\rm{obs}}$ is the observed intensity, and $\sigma_{\rm{ISM}}$ is the cross section of ISM particles.

For the optical and UV bands, we make use of the linear relation between the hydrogen column density and the $V$ band reddening ($A_V$) as presented in \citet{Guver2009}. We also assume $R_V=3.1$. From the reddening curves presented in \citet{Cardelli1989}, we compute the reddening in each band [$A(\lambda)$] and reduce the band luminosity accordingly.

\section{Additional Figures} \label{sec:appC}
Here we show the dynamics and large scale features of models \texttt{s09} and \texttt{m00}. In addition we show detection limits for all four models. See the text for a full description.

\begin{figure}
	\includegraphics[width=\columnwidth]{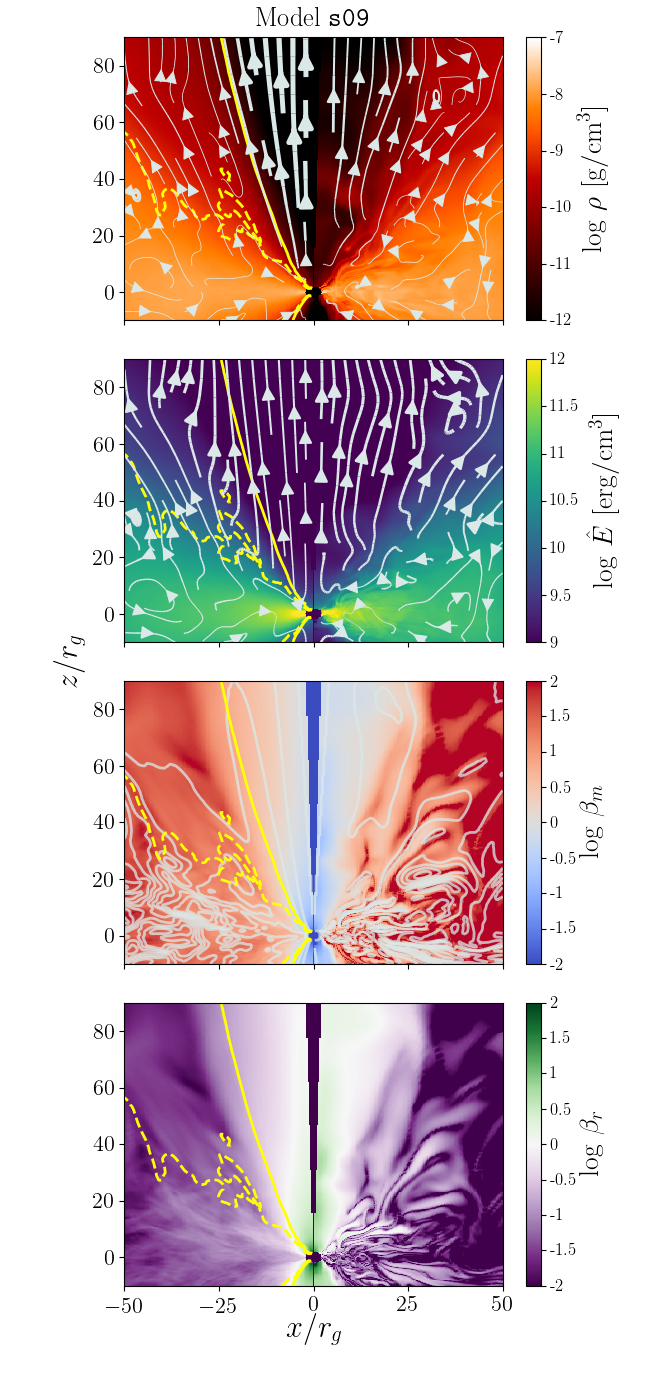}
    \caption{Fluid properties time averaged over $t=15,000-20,000\,t_g$ (left) and for the snapshot at $t=20,000\,t_g$ (right) for the SANE accretion disk model \texttt{s09}. We show gas density with fluid velocity streamlines (top panel), radiation energy density with radiation flux streamlines (second panel), magnetic pressure ratio $\beta_m$ with magnetic field contours (third panel), and radiation pressure ratio $\beta_r$ (bottom panel). The yellow contours in each panel mark the jet/wind boundary (Be=Be$_{\rm{crit}}$, solid yellow) and the wind/disk boundary (Be=$0$, dashed yellow). See the text for a detailed description.}
    \label{fig:figC1}
\end{figure}

\begin{figure}
	\includegraphics[width=\columnwidth]{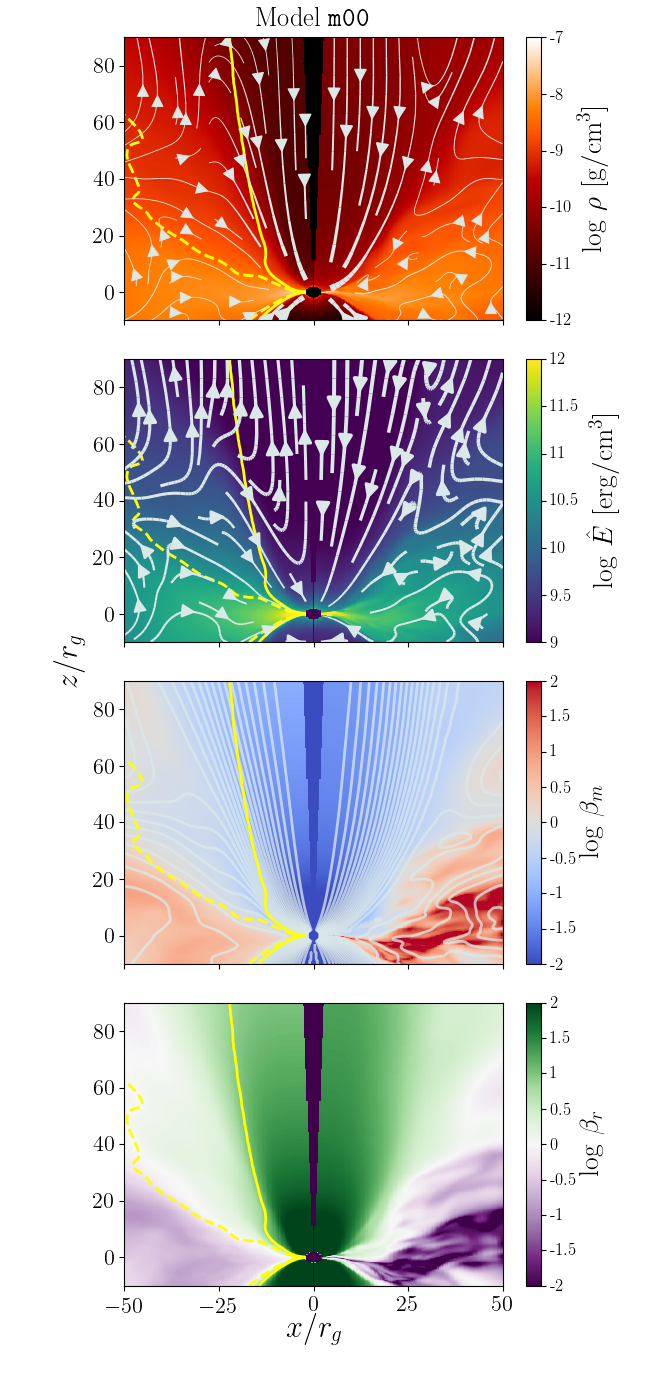}
    \caption{Fluid properties time averaged over $t=20,000-25,000\,t_g$ (left) and for the snapshot at $t=25,000\,t_g$ (right) for the MAD accretion disk model \texttt{m00}. We show gas density with fluid velocity streamlines (top panel), radiation energy density with radiation flux streamlines (second panel), magnetic pressure ratio $\beta_m$ with magnetic field contours (third panel), and radiation pressure ratio $\beta_r$ (bottom panel). The yellow contours in each panel mark the jet/wind boundary (Be=Be$_{\rm{crit}}$, solid yellow) and the wind/disk boundary (Be=$0$, dashed yellow). See the text for a detailed description.}
    \label{fig:figC2}
\end{figure}

\begin{figure}
	\includegraphics[width=\columnwidth]{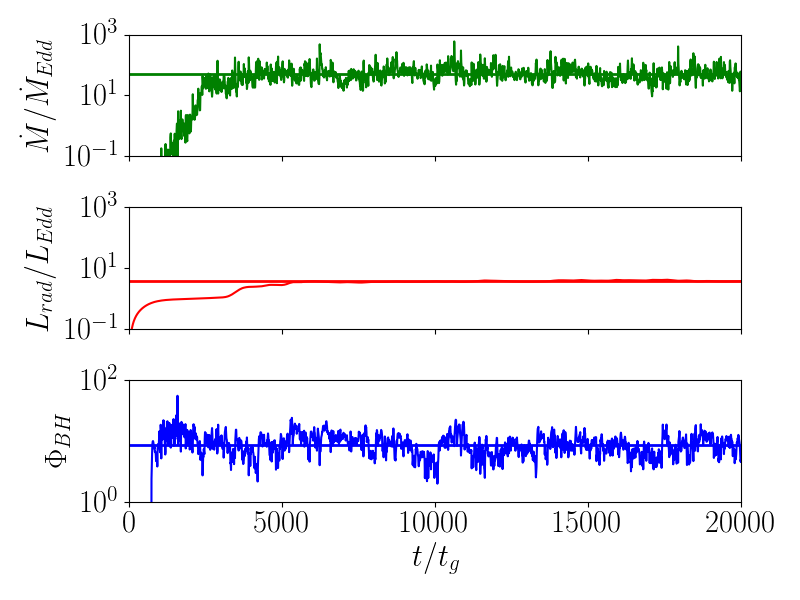}
    \caption{Mass accretion rate (top), radiative luminosity (middle), and magnetic flux parameter $\Phi_{\rm{BH}}$ (bottom) over $t=0-20,000\,t_g$ for the SANE accretion disk model \texttt{s09}. The solid lines show quantities averaged over the last $5000\,t_g$ of the simulation. The disk is evidently SANE for the entire simulation.}
    \label{fig:figC3}
\end{figure}

\begin{figure}
	\includegraphics[width=\columnwidth]{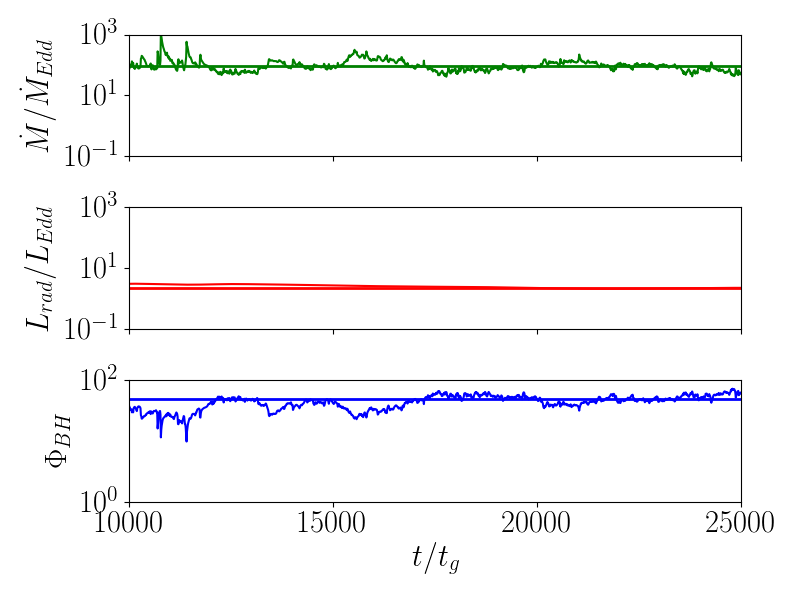}
    \caption{Mass accretion rate (top), radiative luminosity (middle), and magnetic flux parameter $\Phi_{\rm{BH}}$ (bottom) over $t=10000-25,000\,t_g$ for the MAD accretion disk model \texttt{m00}. We only show the data from $t=10000\,t_g$ on since this is after we re-grid from 2D to 3D and perturb the disk. The solid lines show quantities averaged over the last $5000\,t_g$ of the simulation. The disk is evidently in the MAD state.}
    \label{fig:figC4}
\end{figure}

\begin{figure*}
	\includegraphics[width=\textwidth]{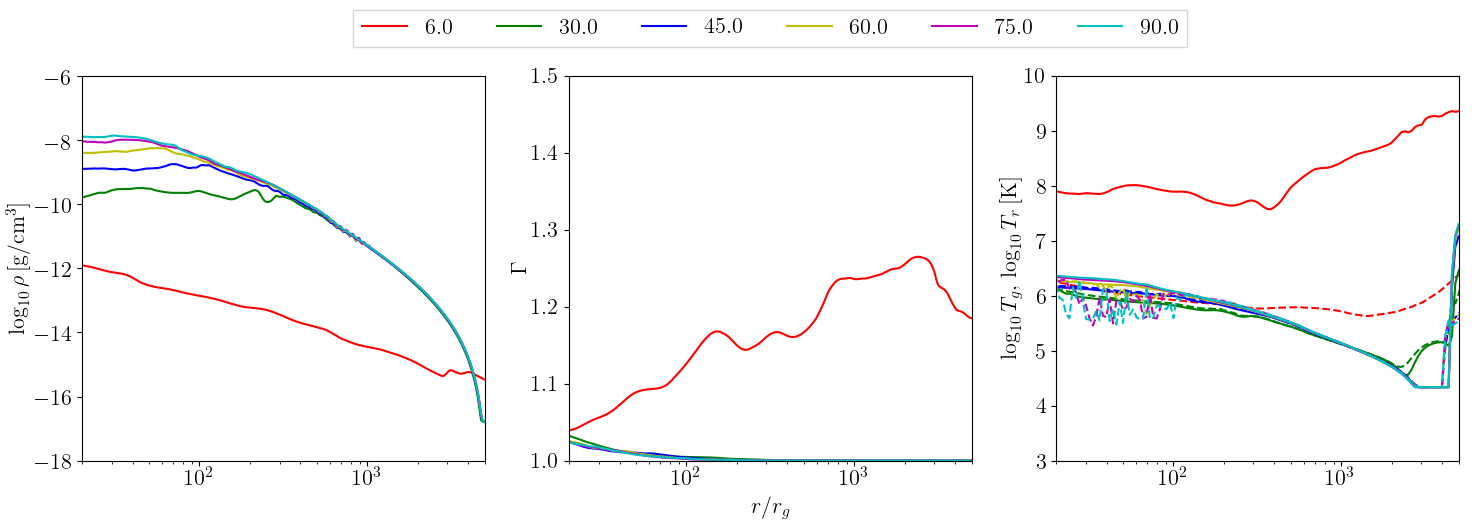}
    \caption{Radial profiles taken at various angles of inclination of gas density (left), Lorentz factor $\Gamma$ (middle), and both gas temperature (solid line) and radiation temperature (dashed line) on the right for the SANE accretion disk model \texttt{s09}.}
    \label{fig:figC5}
\end{figure*}

\begin{figure*}
	\includegraphics[width=\textwidth]{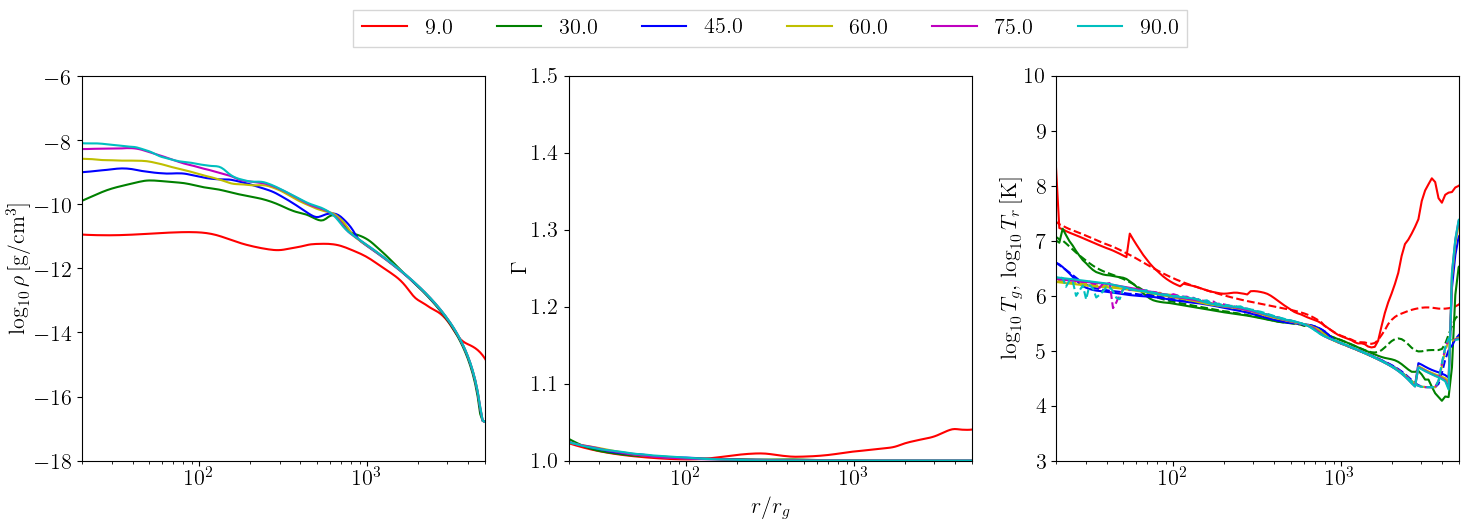}
    \caption{Radial profiles taken at various angles of inclination of gas density (left), Lorentz factor $\Gamma$ (middle), and both gas temperature (solid line) and radiation temperature (dashed line) on the right for the MAD accretion disk model \texttt{m00}.}
    \label{fig:figC6}
\end{figure*}

\begin{figure*}
	\includegraphics[width=\textwidth]{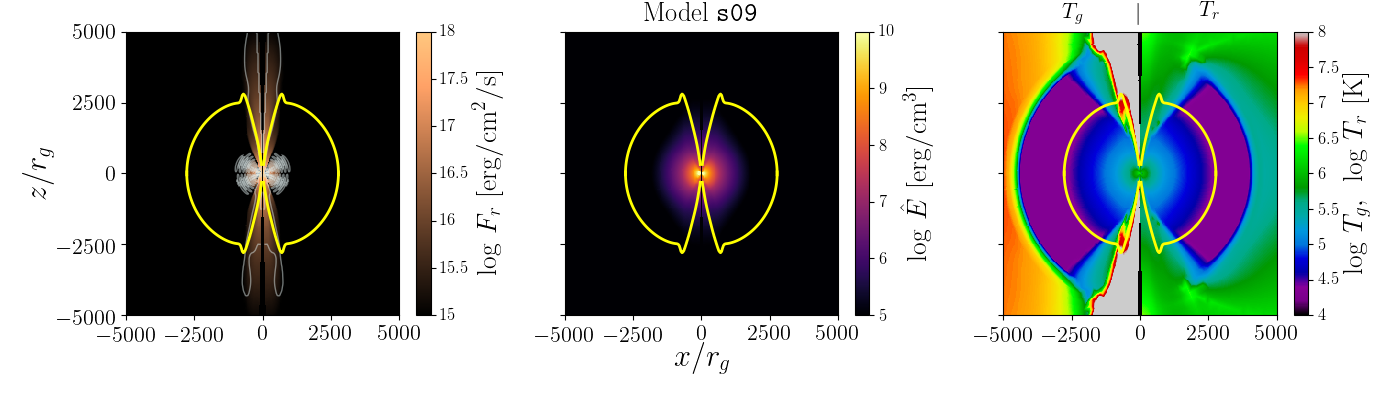}
	\includegraphics[width=\textwidth]{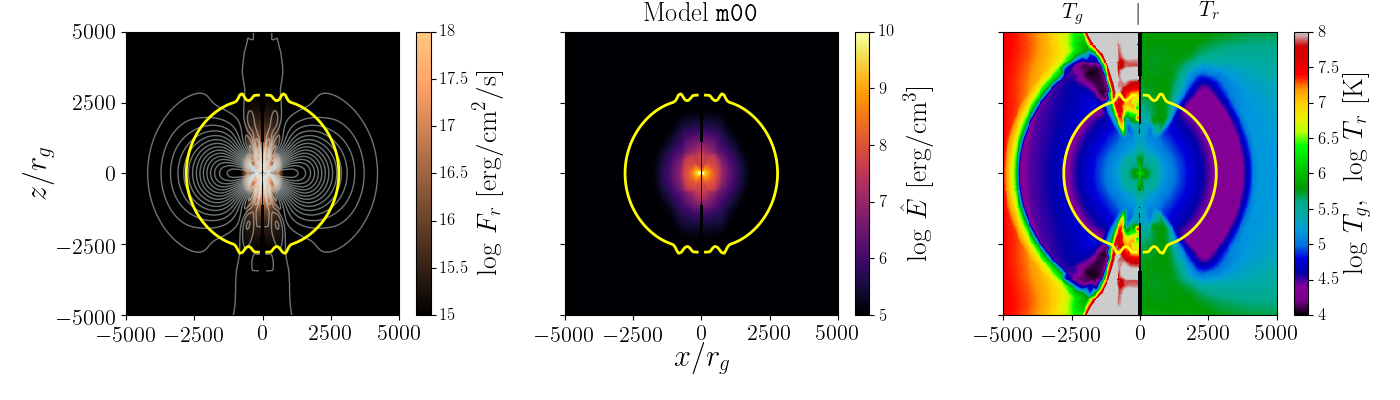}
    \caption{Large scale characteristics of models \texttt{s09} (top) and \texttt{m00} (bottom). In each panel, the yellow contour shows the electron scattering photosphere. (a) The left most panels show the radial flux of radiation radiation (color scale) and vector potential ($A_\phi$, white contours). (b) The middle panel shows the radiation energy density. (c) On the right, we show the radiation temperature.}
    \label{fig:figC7}
\end{figure*}

\begin{figure}
   	\includegraphics[width=\columnwidth]{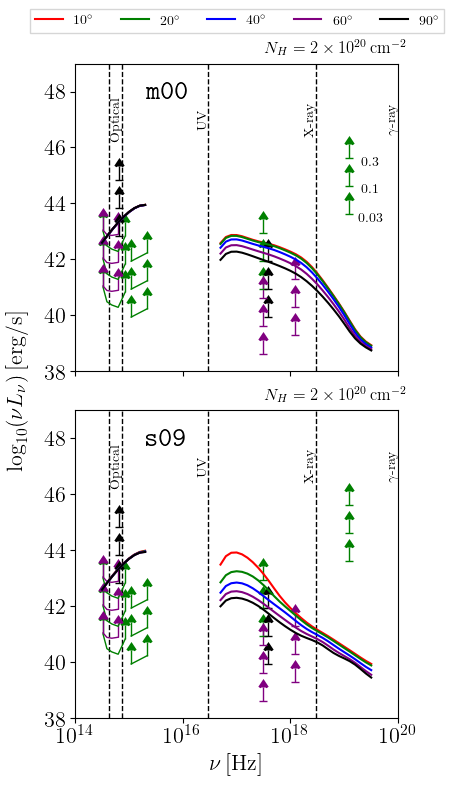}
    \caption{The same as Figure \ref{fig:fig15} but for \texttt{m00} and \texttt{s09}.}
    \label{fig:figC8}
\end{figure}

\begin{figure}
   	\includegraphics[width=\columnwidth]{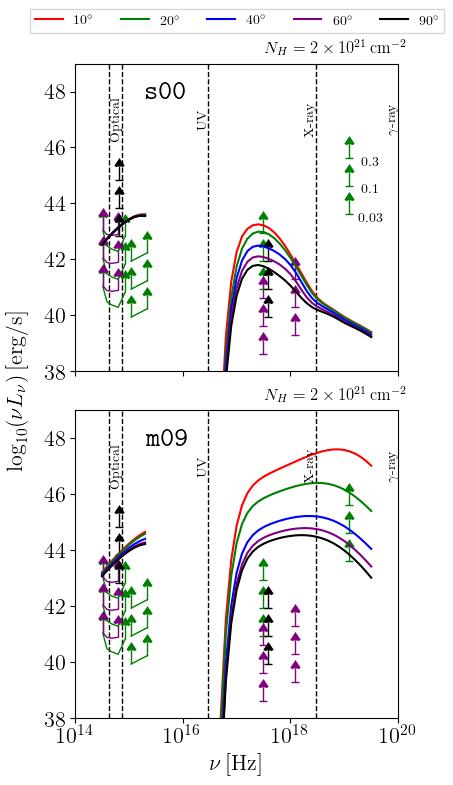}
    \caption{The same as Figure \ref{fig:fig15} but for \texttt{s00} and \texttt{m09} with $N_H = 2\times10^{21}$ $\rm{cm^{-2}}$.}
    \label{fig:figC9}
\end{figure}

\begin{figure}
   	\includegraphics[width=\columnwidth]{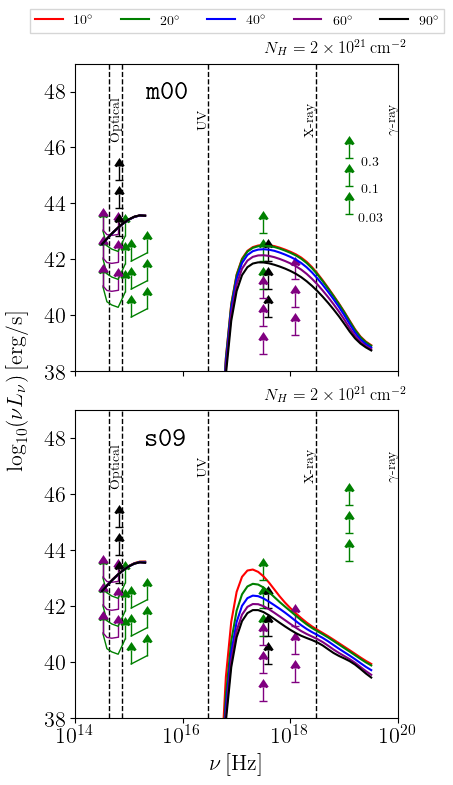}
    \caption{The same as Figure \ref{fig:fig15} but for \texttt{m00} and \texttt{s09} with $N_H = 2\times10^{21}$ $\rm{cm^{-2}}$.}
    \label{fig:figC10}
\end{figure}


\bsp	
\label{lastpage}

\begin{thebibliography}{99}

\bibitem[\protect\citeauthoryear{Abramowicz et al.}{1988}]{Abramowicz1988} 
Abramowicz M.~A., Czerny B., Lasota J.~P., Szuszkiewicz E., 1988, ApJ, 332, 646 

\bibitem[Aleksi{\'c} et al.(2013)]{Aleksic2013} Aleksi{\'c}, J., Antonelli, 
L.~A., Antoranz, P., et al.\ 2013, \aap, 552, A112 

\bibitem[Alexander et al.(2016)]{Alexander2016} Alexander, K.~D., Berger, 
E., Guillochon, J., Zauderer, B.~A., 
\& Williams, P.~K.~G.\ 2016, \apjl, 819, L25 

\bibitem[Aliu et al.(2011)]{Aliu2011} Aliu, E., Arlen, T., Aune, T., et 
al.\ 2011, \apjl, 738, L30 

\bibitem[Auchettl et al.(2017)]{Auchettl2017} Auchettl, K., Guillochon, J., 
\& Ramirez-Ruiz, E.\ 2017, \apj, 838, 149 

\bibitem[Balbus \& Hawley(1991)]{Balbus1991}
Balbus, S.~A., \& Hawley, J.~F.\ 1991, ApJ, 376, 214

\bibitem[Barthelmy et al.(2005)]{Barthelmy2005} Barthelmy, S. D., et al.\ 2005, Space Science Reviews, 120, 143

\bibitem[Berger et al.(2012)]{Berger2012} Berger, E., Zauderer, A., Pooley, 
G.~G., Soderberg, A.~M., Sari, R., Brunthaler, A., 
\& Bietenholz, M.~F.\ 2012, \apj, 748, 36 

\bibitem[Bisnovatyi-Kogan \& Ruzmaikin(1974)]{BisnovatyiKogan1974}
Bisnovatyi-Kogan, G.~S., \& Ruzmaikin, A.~A.\ 1974, Ap\&SS, 28, 45

\bibitem[Blandford 
\& Znajek(1977)]{Blandford1977} Blandford, R.~D., \& Znajek, R.~L.\ 1977, \mnras, 179, 433 

\bibitem[Bloom et al.(2011)]{Bloom2011} Bloom, J.~S., Giannios, D., 
Metzger, B.~D., et al.\ 2011, Science, 333, 203 

\bibitem[Bonnerot et al.(2016a)]{Bonnerot2016a} Bonnerot, C., Rossi, E.~M., 
Lodato, G., \& Price, D.~J.\ 2016a, \mnras, 455, 2253 

\bibitem[Bonnerot et al.(2016b)]{Bonnerot2016b} Bonnerot, C., Rossi, E.~M., 
\& Lodato, G.\ 2016b, \mnras, 458, 3324 

\bibitem[Bonnerot et al.(2017)]{Bonnerot2017} Bonnerot, C., Price, D.~J., 
Lodato, G., \& Rossi, E.~M.\ 2017, \mnras, 469, 4879 

\bibitem[Brown et al.(2015)]{Brown2015} Brown, G.~C., Levan, A.~J., 
Stanway, E.~R., Tanvir, N.~R., Cenko, S.~B., Berger, E., Chornock, R., 
\& Cucchiaria, A.\ 2015, \mnras, 452, 4297 

\bibitem[Burke et al.(1997)]{Burke1997} Burke, B. E., Gregory, J., Bautz, M. W., Prigozhin, G. Y., Kissel, S. E., Kosicki, B. N., Loomis, A. H., \& Young, D. J.\ 1997, IEEE Trans. Electron Devices, 44, 1633

\bibitem[Burrows et al.(2005)]{Burrows2005} Burrows, D.~N., et al.\ 2005, Space Science Reviews, 120, 165

\bibitem[Burrows et al.(2011)]{Burrows2011} Burrows, D.~N., Kennea, J.~A., 
Ghisellini, G., et al.\ 2011, \nat, 476, 421 

\bibitem[Cannizzo et al.(1990)]{Cannizzo1990} Cannizzo, J.~K., Lee, H.~M., 
\& Goodman, J.\ 1990, \apj, 351, 38 

\bibitem[Cardelli et al.(1989)]{Cardelli1989} Cardelli, J.~A., Clayton, 
G.~C., \& Mathis, J.~S.\ 1989, \apj, 345, 245 

\bibitem[Cenko et al.(2012)]{Cenko2012} Cenko, S.~B., Krimm, H.~A., Horesh, 
A., et al.\ 2012, \apj, 753, 77 

\bibitem[Cheng 
\& Bogdanovi{\'c}(2014)]{Cheng2014} Cheng, R.~M., \& Bogdanovi{\'c}, T.\ 2014, \prd, 90, 064020

\bibitem[Coughlin 
\& Begelman(2014)]{Coughlin2014} Coughlin, E.~R., \& Begelman, M.~C.\ 2014, \apj, 781, 82 

\bibitem[Crumley et al.(2016)]{Crumley2016} Crumley, P., Lu, W., Santana, 
R., Hern{\'a}ndez, R.~A., Kumar, P., \& Markoff, S.\ 2016, \mnras, 460, 396

\bibitem[Dai et al.(2013)]{Dai2013} Dai, L., Escala, A., 
\& Coppi, P.\ 2013, \apjl, 775, L9 

\bibitem[Dai et al.(2015)]{Dai2015} Dai, L., McKinney, J.~C., 
\& Miller, M.~C.\ 2015, \apjl, 812, L39 

\bibitem[Dai et al.(2018)]{Dai2018} Dai, L., McKinney, J.~C., Roth, N., 
Ramirez-Ruiz, E., \& Miller, M.~C.\ 2018, \apjl, 859, L20 

\bibitem[Dong et al.(2016)]{Dong2016} Dong, S., Shappee, B.~J., Prieto, 
J.~L., et al.\ 2016, Science, 351, 257 

\bibitem[Evans 
\& Kochanek(1989)]{Evans1989} Evans, C.~R., \& Kochanek, C.~S.\ 1989, \apjl, 346, L13 

\bibitem[Feng et al.(2015)]{Feng2015} Feng, H., Ho, L.~C., Kaaret, P., Tao, 
L., Yamaoka, K., Zhang, S., \& Gris{\'e}, F.\ 2015, \apj, 807, 185 

\bibitem[Fragile et al.(2007)]{Fragile2007} Fragile, P.~C., Blaes, O.~M., 
Anninos, P., \& Salmonson, J.~D.\ 2007, \apj, 668, 417 

\bibitem[G{\"u}ver 
\& {\"O}zel(2009)]{Guver2009} G{\"u}ver, T., \& {\"O}zel, F.\ 2009, \mnras, 400, 2050 

\bibitem[Gammie et al.(2003)]{Gammie2003} Gammie, C.~F., McKinney, J.~C., 
\& T{\'o}th, G.\ 2003, \apj, 589, 444 

\bibitem[Gezari et al.(2008)]{Gezari2008} Gezari, S., Basa, S., Martin, 
D.~C., et al.\ 2008, \apj, 676, 944 

\bibitem[Gezari et al.(2006)]{Gezari2006} Gezari, S., Martin, D.~C., 
Milliard, B., et al.\ 2006, \apjl, 653, L25 

\bibitem[Gezari et al.(2009)]{Gezari2009} Gezari, S., Heckman, T., Cenko, 
S.~B., et al.\ 2009, \apj, 698, 1367 

\bibitem[Giannios 
\& Metzger(2011)]{Giannios2011} Giannios, D., \& Metzger, B.~D.\ 2011, \mnras, 416, 2102

\bibitem[Godoy-Rivera et al.(2017)]{Godoy-Rivera2017} Godoy-Rivera, D., 
Stanek, K.~Z., Kochanek, C.~S., et al.\ 2017, \mnras, 466, 1428 

\bibitem[Guillochon et al.(2014)]{Guillochon2014} Guillochon, J., Manukian, 
H., \& Ramirez-Ruiz, E.\ 2014, \apj, 783, 23 

\bibitem[Guillochon 
\& Ramirez-Ruiz(2013)]{Guillochon2013} Guillochon, J., \& Ramirez-Ruiz, E.\ 2013, \apj, 767, 25 

\bibitem[Guillochon 
\& Ramirez-Ruiz(2015)]{Guillochon2015} Guillochon, J., \& Ramirez-Ruiz, E.\ 2015, \apj, 809, 166

\bibitem[Guillochon 
\& McCourt(2017)]{Guillochon2017} Guillochon, J., \& McCourt, M.\ 2017, \apjl, 834, L19 

\bibitem[Hawley et al.(2011)]{Hawley2011} Hawley, J.~F., Guan, X., 
\& Krolik, J.~H.\ 2011, \apj, 738, 84 

\bibitem[Hayasaki et al.(2016)]{Hayasaki2016} Hayasaki, K., Stone, N., 
\& Loeb, A.\ 2016, \mnras, 461, 3760 

\bibitem[Hills(1975)]{Hills1975} Hills, J.~G.\ 1975, \nat, 254, 295 

\bibitem[Holoien et al.(2016b)]{Holoien2016b} Holoien, T.~W.-S., Kochanek, 
C.~S., Prieto, J.~L., et al.\ 2016, \mnras, 463, 3813 

\bibitem[Holoien et al.(2016a)]{Holoien2016a} Holoien, T.~W.-S., Kochanek, 
C.~S., Prieto, J.~L., et al.\ 2016, \mnras, 455, 2918 

\bibitem[Holoien et al.(2014)]{Holoien2014} Holoien, T.~W.-S., Prieto, 
J.~L., Bersier, D., et al.\ 2014, \mnras, 445, 3263 

\bibitem[Igumenshchev et al.(2003)]{Igumenshchev2003}
Igumenshchev, I.~V., Narayan, R., \& Abramowicz, M.~A.\ 2003, ApJ, 592, 1042

\bibitem[Jansen et al.(2001)]{Jansen2001} Jansen, F., Lumb, D., Altieri, B., et al.\ 2001, A{\&}A, 365, L1

\bibitem[Kaiser et al.(2002)]{Kaiser2002} Kaiser, N., et al.\ 2002, SPIE Conf. Ser., 4386, 154

\bibitem[Kara et al.(2018)]{Kara2018} Kara, E., Dai, L., Reynolds, C.~S., 
\& Kallman, T.\ 2018, \mnras, 474, 3593 

\bibitem[Kara et al.(2016)]{Kara2016} Kara, E., Miller, J.~M., Reynolds, 
C., \& Dai, L.\ 2016, \nat, 535, 388 

\bibitem[Kato et al.(2004)]{Kato2004} Kato, Y., Mineshige, S., 
\& Shibata, K.\ 2004, \apj, 605, 307 

\bibitem[Kawashima et al.(2012)]{Kawashima2012} Kawashima, T., Ohsuga, K., 
Mineshige, S., Yoshida, T., Heinzeller, D., 
\& Matsumoto, R.\ 2012, \apj, 752, 18 

\bibitem[Kelley et al.(2014)]{Kelley2014} Kelley, L.~Z., Tchekhovskoy, A., 
\& Narayan, R.\ 2014, \mnras, 445, 3919 

\bibitem[Kesden(2012a)]{Kesden2012a} Kesden, M.\ 2012a, \prd, 85, 024037

\bibitem[Kesden(2012b)]{Kesden2012b} Kesden, M.\ 2012b, \prd, 86, 064026 

\bibitem[Komossa(2015)]{Komossa2015} Komossa, S.\ 2015, Journal of High 
Energy Astrophysics, 7, 148 

\bibitem[Komossa et al.(2008)]{Komossa2008} Komossa, S., Zhou, H., Wang, 
T., et al.\ 2008, \apjl, 678, L13 

\bibitem[Kroupa et al.(1993)]{Kroupa1993} Kroupa, P., Tout, C.~A., 
\& Gilmore, G.\ 1993, \mnras, 262, 545

\bibitem[Leloudas et al.(2016)]{Leloudas2016} Leloudas, G., Fraser, M., 
Stone, N.~C., et al.\ 2016, Nature Astronomy, 1, 0002 

\bibitem[Levan et al.(2011)]{Levan2011} Levan, A.~J., Tanvir, N.~R., Cenko, 
S.~B., et al.\ 2011, Science, 333, 199 

\bibitem[Levermore(1984)]{Levermore1984}
Levermore, C.~D.\ 1984, J. Quant. Spectrosc. Radiat. Transfer, 31, 149

\bibitem[Lin et al.(2015)]{Lin2015} Lin, D., Maksym, P.~W., Irwin, J.~A., 
Komossa, S., Webb, N.~A., Godet, O., Barret, D., Grupe, D., 
\& Gwyn, S.~D.~J.\ 2015, \apj, 811, 43 

\bibitem[Liu et al.(2015)]{Liu2015} Liu, D., Pe'er, A., 
\& Loeb, A.\ 2015, \apj, 798, 13 

\bibitem[Lodato et al.(2009)]{Lodato2009} Lodato, G., King, A.~R., 
\& Pringle, J.~E.\ 2009, \mnras, 392, 332 

\bibitem[Lodato 
\& Rossi(2011)]{Lodato2011} Lodato, G., \& Rossi, E.~M.\ 2011, \mnras, 410, 359

\bibitem[Lu et al.(2017)]{Lu2017} Lu, W., Krolik, J., Crumley, P., 
\& Kumar, P.\ 2017, \mnras, 471, 1141 

\bibitem[Martin et al.(2005)]{Martin2005} Martin, D. C., et al.\ 2005, Astrophys. J., 619, L1

\bibitem[Metzger et al.(2012)]{Metzger2012} Metzger, B.~D., Giannios, D., 
\& Mimica, P.\ 2012, \mnras, 420, 3528 

\bibitem[Mimica et al.(2015)]{Mimica2015} Mimica, P., Giannios, D., 
Metzger, B.~D., \& Aloy, M.~A.\ 2015, \mnras, 450, 2824 

\bibitem[Mockler et al.(2018)]{Mockler2018} Mockler, B., Guillochon, J., 
\& Ramirez-Ruiz, E.\ 2018, arXiv:1801.08221 

\bibitem[Narayan et al.(2003)]{Narayan2003} Narayan, R., Igumenshchev, 
I.~V., \& Abramowicz, M.~A.\ 2003, \pasj, 55, L69 

\bibitem[Narayan et al.(2012)]{Narayan2012} Narayan, R., S{\"A} dowski, A., 
Penna, R.~F., \& Kulkarni, A.~K.\ 2012, \mnras, 426, 3241 

\bibitem[Narayan et al.(2016)]{Narayan2016} Narayan, R., Zhu, Y., Psaltis, 
D., \& Sa{\c d}owski, A.\ 2016, \mnras, 457, 608 

\bibitem[Narayan et al.(2017)]{Narayan2017} Narayan, R., Sa{\c d}owski, A., \& Soria, R.\ 2017, \mnras, 469, 2997 

\bibitem[O'Riordan et al.(2017)]{Riordan2017} O'Riordan, M., Pe'er, A., \& McKinney, J.~C.\ 2017, ApJ, 843, 81

\bibitem[Novikov 
\& Thorne(1973)]{Novikov1973} Novikov, I.~D., \& Thorne, K.~S.\ 1973, Black Holes (Les Astres Occlus), 343

\bibitem[Paczy{\'n}sky 
\& Wiita(1980)]{Paczynsky1980} Paczy{\'n}sky, B., \& Wiita, P.~J.\ 1980, \aap, 88, 23 

\bibitem[Phinney(1989)]{Phinney1989} Phinney, E.~S.\ 1989, The Center of 
the Galaxy, 136, 543 

\bibitem[Piran et al.(2015)]{Piran2015} Piran, T., S{\c a}dowski, A., 
\& Tchekhovskoy, A.\ 2015, \mnras, 453, 157

\bibitem[Pushkarev et al.(2009)]{Pushkarev2009} Pushkarev, A.~B., Kovalev, 
Y.~Y., Lister, M.~L., \& Savolainen, T.\ 2009, \aap, 507, L33 

\bibitem[Ramirez-Ruiz 
\& Rosswog(2009)]{Ramirez-Ruiz2009} Ramirez-Ruiz, E., \& Rosswog, S.\ 2009, \apjl, 697, L77

\bibitem[Rees(1988)]{Rees1988} Rees, M.~J.\ 1988, \nat, 333, 523 

\bibitem[Roth et al.(2016)]{Roth2016} Roth, N., Kasen, D., Guillochon, J., 
\& Ramirez-Ruiz, E.\ 2016, \apj, 827, 3 

\bibitem[S{\c a}dowski et al.(2014)]{Sadowski2014} S{\c a}dowski, A., 
Narayan, R., McKinney, J.~C., \& Tchekhovskoy, A.\ 2014, \mnras, 439, 503 

\bibitem[S{\c a}dowski et al.(2013b)]{Sadowski2013b} S{\c a}dowski, A., 
Narayan, R., Penna, R., \& Zhu, Y.\ 2013b, \mnras, 436, 3856 

\bibitem[S{\c a}dowski et al.(2015)]{Sadowski2015} S{\c a}dowski, A., 
Narayan, R., Tchekhovskoy, A., Abarca, D., Zhu, Y., 
\& McKinney, J.~C.\ 2015, \mnras, 447, 49 

\bibitem[S{\c a}dowski \& Narayan(2015b)]{Sadowski2015b} S{\c a}dowski, A., 
Narayan, R.\ 2015, \mnras, 453, 2372

\bibitem[S{\c a}dowski et al.(2013a)]{Sadowski2013a} S{\c a}dowski, A., 
Narayan, R., Tchekhovskoy, A., \& Zhu, Y.\ 2013a, \mnras, 429, 3533 

\bibitem[S{\c a}dowski et al.(2016a)]{Sadowski2016a} S{\c a}dowski, A., 
Tejeda, E., Gafton, E., Rosswog, S., \& Abarca, D.\ 2016a, \mnras, 458, 4250

\bibitem[S{\c a}dowski et al.(2016b)]{Sadowski2016b} S{\c a}dowski, A., 
Lasota, J.-P., Abramowicz, M.~A., \& Narayan, R.\ 2016b, \mnras, 456, 3915 

\bibitem[S{\c a}dowski et al.(2016c)]{Sadowski2016c}
S{\c a}dowski, A., Lasota, J.-P., Abramowicz, M.~A., \& Narayan, R.\ 2016c, MNRAS, 456, 3915

\bibitem[S{\c a}dowski et al.(2017)]{Sadowski2017} S{\c a}dowski, A., 
Wielgus, M., Narayan, R., Abarca, D., McKinney, J.~C., 
\& Chael, A.\ 2017, \mnras, 466, 705 

\bibitem[Shappee et al.(2014)]{Shappee2014} Shappee, B. J., Prieto, J. L., et al.\ 2014, ApJ, 788, 48

\bibitem[Shiokawa et al.(2015)]{Shiokawa2015} Shiokawa, H., Krolik, J.~H., 
Cheng, R.~M., Piran, T., \& Noble, S.~C.\ 2015, \apj, 804, 85 

\bibitem[Stone 
\& Loeb(2012)]{Stone2012} Stone, N., \& Loeb, A.\ 2012, Physical Review Letters, 108, 061302

\bibitem[Stone et al.(2013)]{Stone2013} Stone, N., Sari, R., 
\& Loeb, A.\ 2013, \mnras, 435, 1809 

\bibitem[Strubbe 
\& Quataert(2009)]{Strubbe2009} Strubbe, L.~E., \& Quataert, E.\ 2009, \mnras, 400, 2070

\bibitem[\protect\citeauthoryear{Sutherland \& Dopita}{1993}]{Sutherland_1993} Sutherland R.~S., Dopita M.~A., 1993, ApJS, 88, 253

\bibitem[Tchekhovskoy et al.(2012)]{Tchekhovskoy2012} Tchekhovskoy, A., 
McKinney, J.~C., 
\& Narayan, R.\ 2012, Journal of Physics Conference Series, 372, 012040 

\bibitem[Tchekhovskoy et al.(2014)]{Tchekhovskoy2014} Tchekhovskoy, A., 
Metzger, B.~D., Giannios, D., \& Kelley, L.~Z.\ 2014, \mnras, 437, 2744 

\bibitem[Tchekhovskoy et al.(2011)]{Tchekhovskoy2011} Tchekhovskoy, A., 
Narayan, R., \& McKinney, J.~C.\ 2011, \mnras, 418, L79 

\bibitem[Ulmer(1999)]{Ulmer1999} Ulmer, A.\ 1999, \apj, 514, 180 

\bibitem[van Velzen et al.(2011)]{vanVelzen2011} van Velzen, S., Farrar, 
G.~R., Gezari, S., Morrell, N., Zaritsky, D., {\"O}stman, L., Smith, M., 
Gelfand, J., \& Drake, A.~J.\ 2011, \apj, 741, 73 

\bibitem[Wang et al.(2014)]{Wang2014} Wang, J.-Z., Lei, W.-H., Wang, D.-X., 
Zou, Y.-C., Zhang, B., Gao, H., \& Huang, C.-Y.\ 2014, \apj, 788, 32 

\bibitem[Wilms et al.(2000)]{Wilms2000} Wilms, J., Allen, A., 
\& McCray, R.\ 2000, \apj, 542, 914 

\bibitem[York et al.(2000)]{York2000} York, P. J., et al.\ 2000, Astron. J., 120, 1579

\bibitem[Zauderer et al.(2011)]{Zauderer2011} Zauderer, B.~A., Berger, E., 
Soderberg, A.~M., et al.\ 2011, \nat, 476, 425 

\bibitem[Zauderer et al.(2013)]{Zauderer2013} Zauderer, B.~A., Berger, E., 
Margutti, R., Pooley, G.~G., Sari, R., Soderberg, A.~M., Brunthaler, A., 
\& Bietenholz, M.~F.\ 2013, \apj, 767, 152 

\bibitem[Zhu et al.(2015)]{Zhu2015} Zhu, Y., Narayan, R., Sadowski, A., 
\& Psaltis, D.\ 2015, \mnras, 451, 1661 

\end{thebibliography}
\end{document}